\documentclass[twocolumn, aps, prr, superscriptaddress,nofootinbib, sortedaddress,longbibliography, notitlepage, floatfix]{revtex4-2}

\usepackage{ae}
\usepackage[T1]{fontenc}
\usepackage[ansinew]{inputenc}
\usepackage{graphicx, color, soul}
\usepackage[usenames,dvipsnames, pdftex]{xcolor}
\usepackage{bm}
\usepackage{dcolumn}
\usepackage{amsmath,amssymb,amsfonts, mathtools} 
\usepackage[mathscr]{eucal}
\usepackage{textcomp}
\usepackage{appendix}

\usepackage[colorlinks, breaklinks, 
colorlinks=true,
citecolor=Blue,
linkcolor=Blue,
anchorcolor=blue,
urlcolor=Blue,unicode=false]{hyperref}
            
\usepackage[all]{hypcap} 
\usepackage{enumitem}
\usepackage[normalem]{ulem} 
\usepackage{comment}
\usepackage{cancel}

\newcommand{\Le}{\left}
\newcommand{\Ri}{\right}
\newcommand{\nn}{\nonumber}
\newcommand{\f}{\frac}

\newcommand{\ra}{\rangle}

\newcommand{\eq}[1]{\begin{align}#1\end{align}}

\newcommand{\msr}{\mathscr}

\def\Equal{{=}}

\newcommand{\Nc}{$N_\text{csh}$}
\newcommand{\Nred}{$N_\text{sh}$}
\newcommand{\Ndim}{$N_\text{FS}$}
\newcommand{\Isat}{$I_\text{sat.}\,$}

\newcommand{\ketPsiin}{{|\Psi(0)\rangle}}
\newcommand{\braPsiin}{{\langle\Psi(0)|}}
\newcommand{\DeltaP}{\Delta_\mathcal{P}}
\newcommand{\tDeltaP}{\tilde\Delta_\mathcal{P}}
\newcommand{\thop}{{t_\text{hop}}}
\newcommand{\Wfrag}{W_\text{frag}}

\newcommand{\varN}{ \text{var}(N_\text{csh}/N_\text{sh}) }

\begin{document}

\title{Fock space fragmentation in quenches of disordered interacting fermions}

\author{Ishita Modak}
\affiliation{Department of Physics, Indian Institute of Technology Bombay, Mumbai 400076, India}

\author{Rajesh Narayanan}
\affiliation{Department of Physics, Indian Institute of Technology Madras, Chennai 600036,  India}

\author{Ferdinand Evers}
\affiliation{Institute of Theoretical Physics 
and
Halle-Berlin-Regensburg Cluster of Excellence CCE, University of Regensburg, D-93053 Regensburg, Germany}

\author{Soumya Bera}
\affiliation{Department of Physics, Indian Institute of Technology Bombay, Mumbai 400076, India}

\date{\today}

\begin{abstract}
Fock space fragmentation, as it is currently investigated, primarily originates from specific kinematic constraints or emergent conservation laws in many-body systems with translation invariance.  
It leads to non-ergodic dynamics and possible breakdown of the eigenstate thermalization hypothesis. 
Here, we demonstrate that also in disordered systems, such as the XXZ model with random on-site fields, fragmentation appears as a natural concept offering fresh perspectives, for example, on many-body localization~(MBL). 
Specifically, for a given quenched initial state we identify a potential-energy shell in Fock space, defined by its energy and energy uncertainty, which represents the energetically accessible part of the many-body configuration space.
For example, the drastic sample-to-sample fluctuations observed in the weak disorder regime, $W<\Wfrag$, represent fluctuations of the mass of the shell. 
Upon increasing disorder, the potential-energy shell becomes increasingly fragmented.
Its finite-size evolution exhibits a turn-around flow;
extrapolating this flow to the disorder at which the turn-around vanishes
provides a heuristic estimate of the strong-fragmentation scale $\Wfrag$.
We conjecture that the fragmentation identified here provides a geometric mechanism contributing to the finite-time slowing down of relaxation observed in studies of disordered interacting systems.

\end{abstract}

\maketitle

\emph{Introduction}:~The time evolution of charge imbalance is widely used as an indicator to probe thermalization in isolated disordered interacting systems, experimentally and numerically, e.g., see Ref.~\cite{Schreiber2015, AbaninBloch-Review-2018, AletReview2018, SierantReview_2025}.
Starting from a product state such as the N\'eel state, thermalization 
implies that the imbalance decays to zero at long times; conversely, a finite imbalance at long times signals a breakdown of thermalization, marking the emergence of a nonergodic, i.e., many-body-localized (MBL) phase~\cite{Gornyi2005, Basko2006}.

Despite a significant effort, experimentally and numerically, there is no consensus as to whether or not the MBL phase has indeed been observed in generic models of disordered fermions~\cite{Bera2017, Weiner19, PandaMBL19, SirkerPRL20, SuntasPRB20,  Polkovnikov2021, AbaninAOP21, SierantPRB22, Morningstar2022, LongPRL23,  EversPRB23, ChavezUltraSlow23, ColboisPRB24}. In the absence of rigorous analytical results, it is not clear presently in what sense MBL exists. The main detection challenge is taking the double limit of long observation times and large system sizes, which are currently still out of reach numerically and experimentally.

Still, numerical studies have revealed intriguing transient phenomena. We mention only two: (i) The relaxation behavior of quenches changes from accelerated in time to decelerated once the disorder strength $W$ exceeds a threshold $W_c$ that is comparable to the band width of the non-interacting reference system~\cite{Weiner19}. (ii) The sample-to-sample fluctuations in the imbalance relaxation reflect this qualitative change. For a given disorder strength $W\lesssim W_c$, the ensemble consisting of all samples with size $L$ contains sizable fractions of perfectly thermalizing as well as highly insulating samples~\cite{PrelovsekPRB16,Khemani2017,Doggen2018, NandyPRB21}. Comparatively, at $W\gtrsim W_c$ the sample-to-sample fluctuations are moderate.

In this work, we identify a geometric mechanism in Fock space that underlies part of the finite-size and finite-time crossover phenomenology observed at intermediate disorder.
To this end, we embark on the paradigmatic model of disordered, spinless fermions

\newcommand{\Hprime}{{\hat H^\prime}}
\begin{align}
    \hat{\msr{H}} \coloneqq \hat T + \Hprime 
    \label{e1}
\end{align} 
with $\Hprime \coloneqq \hat V + \hat U$ and 
\begin{align} 
    \hat T \coloneqq & -\frac{\thop}{2}\sum_{x=1}^{L-1}(\hat{c}_{x}^{\dagger} \hat{c}_{x+1} +\text{h.c}) \nn \\
       \hat U \coloneqq & \sum_{x=1}^{L} \epsilon_x \! \Le(\hat{n}_{x}{-}\f{1}{2}\Ri) \nn \\
       \hat V \coloneqq & V \sum_{x=1}^{L-1}\Le(\hat{n}_{x}{-}\f{1}{2}\Ri)  \Le(\hat{n}_{x+1}{-}\f{1}{2}\Ri); \nn
\end{align}
here, $x$ labels the lattice sites, $L$ denotes the systems size, and $\hat{n}_x {=} \hat{c}_x^{\dagger}\hat{c}_x$ is the local density operator. We choose interaction $V{=}\thop$ throughout, and $\epsilon_x$ denotes the onsite random potential drawn independently from a uniform distribution in $[-W, W]$.

Following previous authors\footnote{
The Fock space perspective has been successfully adopted before, for example, in understanding electron lifetimes in interacting quantum dots by mapping to a single-particle localization problem in Fock-space~\cite {Altshuler1997}.
More recently, significant progress has been made by approximating Fock-space as a Bethe lattice (a tree without loops)~\cite{TikhonovBethePRB16,SonnerPRB17,BiroliPRB22, rizzoBethe24} or by the random regular graph~\cite{GarciaPRl17, BeraRRG18, GarciaMataRRGPRR20, detomasiRRGPRB20, TikhonovAnn21, SierantSci23}.
Also, the MBL transition has been investigated from the Fock-space perspective~\cite{MaceMultifractality2018,RoyPercPRB19a, pietracaprina2019hilbert, detomasiFSPRB21, SutradharPRB22,  ThibaultPRB24, RoyReview_2025}, including certain aspects of dynamics~\cite{CreedPRB23, BiroliPRB24, ScoquartPRB25, SunPRB25}. 
}~(see also Ref.~\cite{SuppMat} for further discussion), we adopt a Fock-space perspective of the many-body dynamics. 
Our focus will be on samples in the regime of moderate to strong disorder, i.e., $W\gtrsim \thop$, comparable to the single-particle band width in Eq.~\eqref{e1}. 
In this regime, the potential energy, 
$\hat H^\prime$, 
begins to dominate and therefore the site-occupation basis is a natural choice for the Fock-space construction: every basis state, $|b\rangle$, is an eigenstate of the local occupations, $\hat n_x$, and hence also an eigenstate of $\hat H^\prime$ 

with corresponding eigenvalue $E_b$. 

The relaxation dynamics we are after is observed in quench protocols performed in isolated finite-size systems. The corresponding time evolution conserves the initial state's energy $E_\text{in}$ and, in fact, all moments of the Hamiltonian $\hat{\msr{H}}$. To the extent that the potential energy dominates, this implies that the time evolution involves all Fock-space sites $|b\rangle$ with energies $E_b$ near $E_\text{in}$ within a window given by the energy variance of the initial state. We refer to this set of Fock-space sites as ``potential-energy shell''; it provides an effective ``dynamically active region" of Fock space.

We will show that the concept of the potential-energy shell in Fock space turns out to be fruitful in multiple ways. 
For example, the sample-to-sample fluctuations at moderate disorder turn out to have a geometric interpretation: they reflect fluctuations of the effective size of the potential-energy shell. 
Moreover, the concept also sheds light on the somewhat counterintuitive observation that sample-to-sample fluctuations appear weaker at stronger disorder. Here is underlying a ``shell fragmentation'': once $W$ increases above two times the bare bandwidth ($W\approx 3.5-4$), the potential-energy shell falls into ``fragments''; the distribution of fragments that host the initial state happens to be narrow (within a window of system sizes) and therefore sample-to-sample fluctuations are suppressed~\cite{SuppMat}.

Fragmentation strongly restricts the directly connected region available to the initial state. Relaxation beyond this region then requires processes that couple different neighboring fragments, leading naturally to slow dynamics.
We stipulate that this slowing down has been observed in many studies, numerical and experimental, as `creep' and gave rise to introducing accelerating and decelerating regions in phase-space in Ref.~\cite{Weiner19}. 

Finally, we present an analysis of the cluster statistics suggesting that, unlike percolation, the fragmentation transition exhibits a regime of turn-around flow whose floating towards strong coupling is just transient; only at very large disorder values, the attractive realm of the strong-coupling fixed point is ultimately reached. 
Relating to earlier work on MBL, an intermediate flow towards the insulator has been reported~\cite{NieddaRG2024};  since the turn-around is seen already at moderate disorder $W\approx 3.5$, we believe that it reflects fragmentation occurring in the numerical reference data. 
We mention that there is growing evidence that also in the MBL context, the strong-coupling fixed point attains stability - if at all - only at extremely strong disorder.

We emphasize that shell fragmentation is a geometric property of the Fock-space graph and is not, by itself, a diagnostic of MBL. Quantum tunneling between fragments and interference effects remain outside the present construction.

\emph{Method:}~
The relaxation of the initial state is described by the Hamiltonian \eqref{e1}. For time propagation, we rely upon the standard Chebyshev expansion technique; the technical details are given in Ref.~\cite{NandyPRB21, SierantReview_2025}. 

\begin{figure*}[t]
    \centering
    \includegraphics[width=\textwidth]{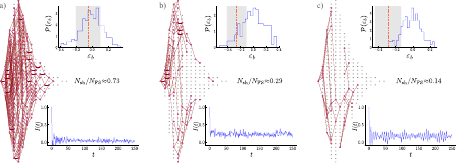}
    \caption{Fock space structure for a system of size $L \Equal 10$, shown for three different disorder realizations at $W \Equal 1.5$. The leftmost site corresponds to the N\'eel state, and the states to the right are ordered by one spin flip distance from it. The vertical arrangement of the sites is arbitrary. 
    The red dots indicate the potential-energy shell; the connectivity of the basis states introduced by the hopping term in Eq.~\eqref{e1} is indicated by lines. 
    Upper inset: The distribution $\mathcal{P}(\varepsilon_b)$ of the Fock-space energy density $\varepsilon_b{=}E_b/L$, where $E_b$ is defined in Eq.~\eqref{e3a}; the red vertical line marks the energy of the N\'eel state, and the shaded region indicates the energy variance, 
    $\Delta_\text{N\'eel}$. Lower inset: Imbalance relaxation $I(t)$ for the corresponding sample. 
    \label{f1}}
\end{figure*}

%
Our primary observable for relaxation processes is the sublattice imbalance,
\eq{
I(t) \coloneqq \frac{2}{L} \sum_{x=1}^L (-1)^x \langle \Psi(t)| \hat{n}_x|\Psi(t)\rangle 
\label{eq:imbalance}
}
where $|\Psi(t)\rangle {\coloneqq} e^{-i \hat{\mathscr{H}}t} \, |\text{N\'eel}\rangle$ with the initial state at $t=0$ being the N\'eel state, represented as $|\text{N\'eel}\rangle \coloneqq |1010\ldots\rangle $ in the site-occupation basis. We fix the particle number to be $L/2$, which defines the dimension of the FS \Ndim $\Equal \binom{L}{L/2}$. In the supplementary material~\cite {SuppMat}, we tested the sensitivity to initial conditions by starting from other initial states.

\emph{Potential-energy shell concept:}~ 
We represent Fock-space in the occupation basis, $|b\rangle$, which consists of all states that are eigenstates of the occupation number operators $\{\hat n_{x}, x=1,\ldots,L\}$; by construction $\Hprime |b\rangle = E_b|b\rangle$ and $E_b=\langle b| \hat{\msr{H}} |b\rangle$ because 
$\langle b|\hat T|b\rangle=0$. 
The eigenvalues are given by 
$
    E_b = \frac{1}{2}\sum_{x=1}^{L} \epsilon_x \! \sigma_x
	+ \frac{V}{4} \sum_{x=1}^{L-1}\sigma_{b,x}\sigma_{b,x+1}$
where $\sigma_{b,x} = 2(n_x-1/2)$ denote a sequence of $L$ ``bits'' characteristic for every basis vector $|b\rangle$. The notation emphasizes the close relation between the statistics of the potential energy content of $|b\rangle$ and the thermodynamics of the 1d-Ising model. 

Towards the flat-band limit, 
$\thop\ll V,W$, one expects that the relaxation dynamics is approximately confined to the subset of all basis states with 
\begin{align}
    |E_\text{in}-E_b|\lesssim \alpha \Delta_\text{in}
    \label{e3a}
\end{align}
where 
$\Delta_\text{in} \coloneqq (\braPsiin \hat{\msr{H}}^2-E_\text{in}^2\ketPsiin)^{1/2}$ and $\alpha$ a phenomenological parameter of order unity,
which we take as $\alpha{=}1/2$. 
The parameter $\alpha$ sets the width of the shell; the qualitative fragmentation behavior is robust against moderate variations of $\alpha$, as shown in the Supplementary Material.
The set of all basis states that satisfy the condition \eqref{e3a} we refer to as the potential-energy shell. The number of states contained in this set, \Nred, we refer to as the `mass' of the shell.

In the following, we focus on initial states, $\ketPsiin$, that have been chosen as one of the basis vectors $|b\rangle$, e.g., $\ketPsiin=|\text{N\'eel}\rangle$.  
The corresponding variance $\Delta_\text{in}$ scales with the kinetic energy content of $|\Psi(0)\rangle$. Specifically, for the N\'eel state 
 $\Delta_\text{in}{=}\Delta_\text{N\'eel}$ with $\Delta_\text{N\'eel} =(\thop/2)\sqrt{L-1}$; we notice that 
 $\Delta_\text{N\'eel}$ represents an upper boundary for the kinetic-energy content of all basis states, 
 $\Delta_\text{N\'eel}\geq \langle b|\hat T^2|b\rangle^{1/2}$.

\emph{Illustration:}~ 
Figure~\ref{f1} illustrates the concept. It displays three examples for potential-energy shells (red circles) embedded into the full Fock space (red and grey circles) taken at moderate disorder $W=1.5$. The lines denote pairs of basis states, 
$|b\rangle, |b'\rangle$, that are linked neighbors in the sense that $\sum_{x=1}^{L-1} \langle b'|\hat c^\dagger_x \hat c_{x+1} + \text{h.c.}|b\rangle >\neq 0$. Two basis states ("sites") are called connected if there is a continuous path along linked neighboring sites between them; two connected sites are in the same cluster. While only a single cluster is seen in Fig. \ref{f1}a, multiple clusters exist in Fig. \ref{f1}b,c.

The potential-energy shells displayed in Fig. \ref{f1} define the active space available when quenching a N\'eel state. Representing the dynamically active part of the Fock space, fluctuations of the potential-energy shell manifest in sample-to-sample fluctuations of the time-evolving quenches. This is why the concept is so interesting. Furthermore, in Ref.~\cite{PrasadPRB22}, the quench dynamics for different initial conditions show a state-dependent imbalance decay, which can now be understood as the presence of distinct dynamically active parts in Fock space.

\begin{figure}[!b]
    \centering
    \includegraphics[width=1\columnwidth]{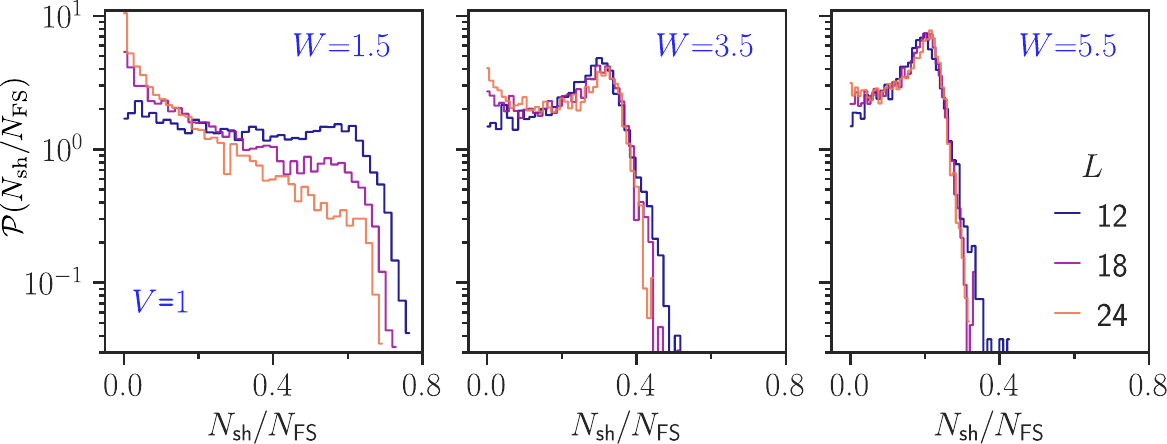}
    \caption{Shows the distribution of the normalized mass of the potential-energy shell, \Nred/\Ndim,  for different $W, L$.  
    }
    \label{f2}
\end{figure}

\emph{Fluctuations of the mass of the potential-energy shell:}~
There are three key observations to be made in Fig.~\ref{f1}:
(i) For the three typical moderate-disorder samples shown, the N\'eel state's energy, $E_\text{in}$, fluctuates strongly with respect to the distribution $\mathcal{P}(E_b)$ of the site energies $E_b$ (upper inset). 
This is qualitatively expected: 
First, we note that $E_\text{in}$ exhibits disorder-induced fluctuations which are of order 
$\sqrt{4/3 (W/4)^2 L}$ with box-distributed disorder.  Second, within a Gaussian model, one expects 
$\mathcal{P}(E_b)=(2\pi\DeltaP^2)^{-1/2}
e^{-E_b^2/2\DeltaP^2}$ with 
$
\DeltaP=\sqrt{
(V/4)^2(L-1) +4/3(W/4)^2 L}
$.~\footnote{The variance of the on-site energies does not incorporate a term resulting from the kinetic energy because it reflects the distribution of eigenvalues of $\Hprime$; indeed, the variance of the on-site energies, 
$\langle b|\hat{\msr{H}}^2-E_b^2|b\rangle$, averaged over the Fock space is given  by $\DeltaP^2+\frac{1}{2}\Delta_\text{N\'eel}^2$.}
Since, third, the disorder average of 
$\overline{E_\text{in}}$ is typically extensive, 
e.g., for the N\'eel state 
$
\overline{E_\text{N\'eel}}=-(V/4)L$, 
one estimates that the system size shown in Fig.~\ref{f1} is still at the crossover at which $E_\text{in}$ begins to migrate to the tails of $\mathcal{P}$ with increasing $L$.

(ii) The mass of the potential energy shell, i.e., \Nred, is highly sensitive to its average energy. As one would expect, for average energies near the edges of $\mathcal{P}(E_b)$, the shell volume's fraction of the total Fock space \Nred/\Ndim is reduced. Typical factors are 2.5 or even 5 as extracted from the examples displayed in Fig.~\ref{f1}.

Shedding fresh light on the strong fluctuations seen in Fig.~\ref{f1}, we analyze the mass fluctuations of the potential-energy shell. 
Figure~\ref{f2} displays the evolution of the distribution of the shell-mass fraction with disorder strength and system size. Indeed, a broad, almost flat distribution at $W{=}1.5$ explains the very large sample-to-sample fluctuations seen in Fig.~\ref{f1} and reported in the literature in the respective regime of $W$ and $L$~\cite{Doggen2018,  Torres_SelfPRB20, NandyPRB21, EversPRB23}.

\begin{figure}[!t]
    \centering
\includegraphics[width=0.8\columnwidth]{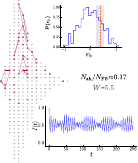}
    \caption{Similar data as in Fig.~\ref{f1} for disorder strength $W{=}5.5$ for $L=10$. With $W$ growing beyond $W\approx 3.5$, the energy shell is seen to become increasingly fragmented. Correspondingly, $I(t)$ does not decay.  The resulting oscillations in $I(t)$, which is a hallmark of fragmentation, are analyzed in detail in Ref.~\cite{SuppMat}(see Fig. S1 therein).  
        }
    \label{f3}
\end{figure}
With respect to the evolution of the distribution with system size, we notice that finite-size effects are very large: characteristic of all $W$-values displayed in Fig.~\ref{f2}, the fraction of samples with $\text{\Nred/\Ndim}\lesssim 0.1$ is increasing rapidly with $L$. We take this as a manifestation of the fact that the overlap of the distributions of $E_b$ and $E_\text{in}$ - estimated by 
$\mathcal{P}(\overline{E_\text{in}})$ -  slowly decreases with increasing $L$; within the Gaussian model it is expected  
$
\mathcal{P}(\overline{E_\text{in}})\approx 
e^{-\frac{1}{2} \mathfrak{r}_\text{in}^2 L }
/\sqrt{2\pi\tDeltaP^2}
$
where $\mathfrak{r}_\text{in} \coloneqq 
(\overline{E_\text{in}}/L)/(\tDeltaP/\sqrt{L})$; here $\tDeltaP^2 = \DeltaP^2 + \Delta_\text{in}^2$. 
Correspondingly, the average fraction \Nred/\Ndim is expected to be suppressed exponentially with increasing $L$. 

Within the Gaussian model, the exponential suppression starts once $L$ exceeds the crossover scale 
$L_\text{W}\coloneqq 2/\mathfrak{r}_\text{in}^2$. In this model, we can give qualitative estimates, e.g., 
$
\mathfrak{r}_\text{in}^{-2}{=}(2\thop/V)^2 {+} 1{+} 4/3 (W/V)^2
$. For the present situation $V=\thop$, we are led to conclude that 
$L_W\approx (10+(8/3)W^2)$
which is roughly consistent with what is seen in Fig.~\ref{f2}. 
Similarly, at $W\approx 3$ the disorder dominates the scale $L_W$, making it large and essentially inaccessible for us, which is also consistent with Fig.~\ref{f2}. 

(iii)  
At system sizes $L\lesssim L_W$, the satellite peak seen in Fig.~\ref{f2} dominates the behavior of typical samples. The preceding argument suggests that this peak gradually disappears in the limit of large $L$. 
However, the satellite dominates the transient dynamics, e.g., the sample-to-sample fluctuations, making them appear {\em smaller} in systems with {\em larger} disorder. More importantly, the satellite peak is a remnant of the non-interacting $V=0$  limit, and the role of the interaction is to facilitate the flow of the weight from the peak to \Nred/\Ndim $\rightarrow 0$~(see also Ref.~\cite{SuppMat}).

\emph{Connectivity fluctuations - fragmentation:}~ 
Figure \ref{f2} implies that within a wide window of sizes $L\lesssim L_W$ a substantial fraction of the system remains part of the percolation cluster, with typical samples exhibiting a mass fraction exceeding \( \text{\Nred}/\text{\Ndim} \gtrsim 20\% \) at disorder values $W\gtrsim 3$.
As is illustrated in Fig.~\ref{f3}, such rather massive potential-energy shells are not necessarily ergodic. The reason is that the shell consists of different parts that may or may not be connected; we refer to them as `fragments'. 
Clearly, for relating the potential energy shell to the relaxation dynamics, the shell geometry is to be carefully characterized.\footnote{ Also in the opposite case, $\text{\Nred}/\text{\Ndim}\ll 1$, there is not necessarily a direct connection between the mass of the shell and the relaxation behavior. Even if the shell has a mass of measure zero, the quench that it carries can be ergodic if the shell is space-filling, i.e., some of its sites can be found in any region of Fock-space. }

\begin{figure} 
    \centering
   \includegraphics[width=1\columnwidth]{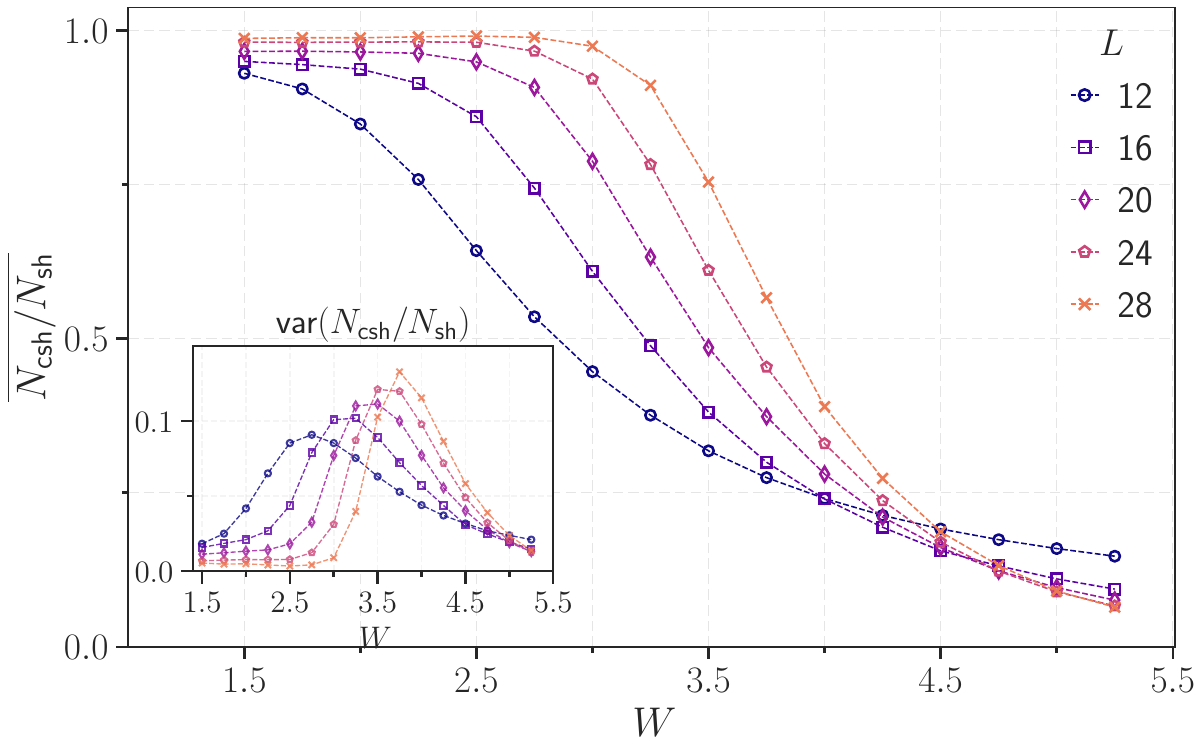}
    \caption{The size of the cluster that contains the initial state, \Nc, normalized to the size of the corresponding potential-energy shell, \Nred, as a function of the disorder strength $W$ for system sizes $L{=}12,\ldots 28$. Inset: The corresponding fluctuations of the ratio \Nc/\Nred. 
    \label{f4}}
\end{figure}

\begin{figure*}[t]
    \centering
    \includegraphics[width=0.95\textwidth]{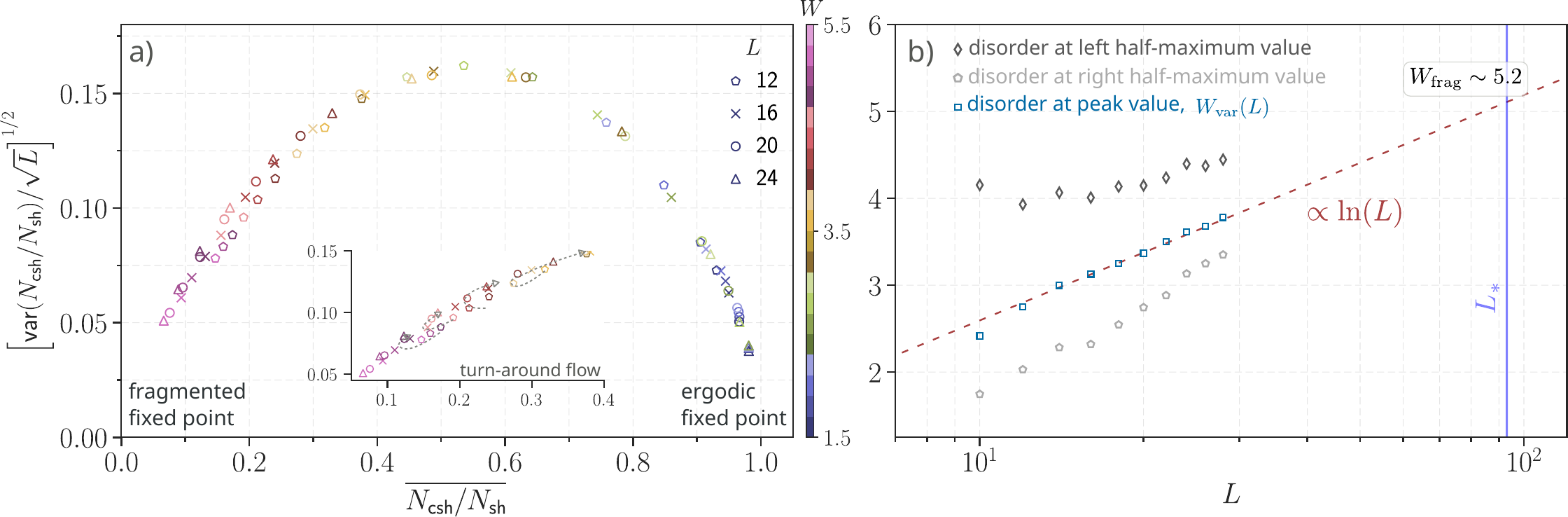}
    \caption{a) Direct comparison of variance and mean values, re-plotting the data Fig.~\ref{f4}. The colors indicate different disorder values $W$ as given in the legend. An approximate collapse is achieved after normalizing the ordinate with the heuristic factor of $\sqrt{L}$. Inset: Zoomed left part of the main plot, showing the turn-around flow. b) Flow of the maximum position $W_\text{var}$ of $\varN$ with $L$, as extracted from data such as shown in inset of Fig.~\ref{f4}. (blue squares) The dashed line indicates a logarithmic extrapolation; the vertical line represents the system size $L_*$ at which the turn-around of the RG-flow from strong to weak coupling is cut off. The other two traces indicate the evolution of the disorder values at which $\varN$ takes half the peak value (left, diamonds and right, pentagons). All traces taken together incorporate the evolution of the peak's shape with system size. 
    }
    \label{f5}
\end{figure*}
In order to detect 
fragmentation, we study in Fig.~\ref{f4} the statistics of the mass of the fragment, \Nc, that contains the initial state $|\text{N\'eel}\rangle$. 
At moderate to weak disorder, in nearly all samples, the potential-energy shell is singly connected, $\overline{ N_\text{csh}/ N_\text{sh} } \approx 1$, corresponding to the connected fixed point.
In particular, as seen in the inset, the variance of \Nc/\Nred is very close to zero at $W\lesssim 3$ once the system size exceeds a value of $L\simeq 24$. 
Fragmentation implies that the initial state is not connected with most of the potential energy shell, i.e. 
$\text{\Nc}/\text{\Nred}\ll 1$. As seen in Fig.~\ref{f4}, this is the dominating behavior at strong disorder, $W\gtrsim 4.5$.

Upon inspecting the evolution of the traces seen in Fig.~\ref{f4}, the question naturally arises if there is a critical disorder, $\Wfrag$, at which phase transition takes place, similar to percolation, which separates a phase dominated by fragmentation from a singly-connected regime. 
The range of system sizes at our disposal is not suitable to provide a definite answer by extrapolating $L\to\infty$. We notice, however, that for a sharp transition to occur, all traces will have to intersect in a single point in the limit of large $L$, which defines the critical disorder $\Wfrag$. 
Since the intersection point of two neighboring traces in Fig.~\ref{f4} 
is seen to move rapidly to larger disorder, a conservative lower bound would be  $\Wfrag\gtrsim 4$ with the distinct possibility that there is no bound at all.

Even for a critical point floating towards infinite disorder, a meaningful definition of a strong-disorder phase can be given within a familiar scenario of turn-around flow~\cite{Cardy96,Cardy99,ZhuWanPRB15}. 
This point has also been emphasized in recent work on the MBL transition~\cite{NieddaRG2024}. 
The basic idea is readily identified in Fig.~\ref{f4}: 
in the range $4{\lesssim}W{\lesssim} 4.5$ the volume fraction 
$\overline{N_\text{csh}/N_\text{sh}}$ is seen to first decrease before it increases again at larger $L$. At even larger $W$, the $L$-window at which $\overline{N_\text{csh}/N_\text{sh}}$ decreases can become big enough so as to be interpreted as the RG-flow approaching a fully fragmented fixed point. The flow reversal at the largest system sizes is then interpreted as instability of the fragmented fixed point.

%
\emph{Evolution of fragmentation with system size:}~ 
In an attempt to settle the question if $\Wfrag$ is finite upon $L\to\infty$, we compare in Fig.~\ref{f5}a)  the average $\overline{N_\text{csh}/N_\text{sh}}$ to the variance for the entire spread of disorder values and system sizes. In the regime of weak to moderate disorder,  
$\text{\Nc}/\text{\Nred}\lesssim 1$, we observe a collapse of the data consistent with  
$
\text{var}(\text{\Nc}/\text{\Nred}) = \sqrt{L} 
f(\overline{\text{\Nc}/\text{\Nred}})
$
where $f(x)$ denotes a parameterless function of its argument. The $\sqrt{L}$ rescaling is empirical and is introduced to organize the finite-size data; we do not interpret it as a controlled scaling theory.
The master curve hosts families of data points that share the same disorder, $W$. With $L$ increasing, the data points per family, flow along the master curve towards the weak-disorder fixed point $\overline{N_\text{csh}/N_\text{sh}}=1$. 

At the strongest disorder, $W>5$, the family flow seen in Fig.~\ref{f5}a) is directed opposite, namely towards the fragmented fixed-point at which \Nc becomes a set of measure zero within the potential-energy shell. 
Crucially, as witnessed in the inset of Fig.~\ref{f5}a), at slightly weaker disorder, a turnaround is visible: the initial flow is directed towards the fragmented fixpoint before it redirects once a system size has been crossed that depends on the disorder strength $L^{*}(W)$. Moreover, it is only at $L>L^{*}$ that the flow returns to the master curve $f(x)$; correspondingly, the data for the regime of strong disorder, $0<\overline{\text{\Nc}/\text{\Nred}}\lesssim 0.4$, collapses only partially onto the master curve, namely after the turn-around.

Being limited to system sizes in our numerics, we can observe the turnaround proper only for disorder values so that $L^{*}(W)\lesssim24$, which implies the upper limit $W\approx 4.6$. The clustering of larger-$W$ data at the largest accessible sizes suggests that the turnaround persists beyond this range.
The collapse in  Fig.~\ref{f5}a) gives
$
\left[\frac{\mathrm{var}(N_{\rm csh}/N_{\rm sh})}{\sqrt{L}}\right]^{1/2}
\lesssim 0.161 .
$
Since $0\leq N_{\rm csh}/N_{\rm sh}\leq1$, the variance is bounded by
$1/4$. If the peak reaches this maximal value, one obtains
$
\left[\frac{1/4}{\sqrt{L_*}}\right]^{1/2}\approx0.161,
$
and hence $L_*\approx93$. 
At system sizes exceeding $L_*$, the RG-flow observed in \ref{f5}a) is confined to regions either left or right of the maximum, without the possibility to cross. This suggests the following scenario that at $L\gtrsim L_*$ all points with 
$\overline{\text{\Nc/\Nred}}<1/2$ 
fall into the domain of attraction of the fragmented fixed point, while the others flow towards ergodicity.

Since $L_*$ is larger than the system sizes reliably accessible in current computational studies, $L\lesssim28$, the extrapolation
necessarily remains heuristic.
Nevertheless, the result is of interest because it permits an extrapolated estimate of $\Wfrag$. 
In Fig.~\ref{f5}(b), we follow the evolution of the maximum position of the variance, $W_\text{var}(L)$, displayed in Fig.~\ref{f4}, inset with system size $L$. It is of interest since one might identify $\Wfrag \approx W_\text{var}(L_*)$. Within the limited set of system sizes available, we have $W_\text{var}(L)\sim \ln(L)$ and the extrapolation to $L_*$ 
yields $\Wfrag \approx 5$.

\emph{Conclusion:}
We analyze quenches in the one-dimensional disordered t-V model~\eqref{e1} within a Fock-space representation. 
At the core of our analysis, we introduce the concept of a potential-energy shell, which essentially contains all Fock-space basis states that are energetically accessible within the relaxation dynamics. Crucially, the shell is a property of the quench protocol, which is defined by the Hamiltonian {\it together} with the initial state, taken to be the N\'eel state in this work and also other initial states~\cite{SuppMat}.  

As one would expect, we observe a close relationship between dynamical observables, such as the sublattice imbalance, and the geometric properties of the shell, i.e., its size (`mass'), connectivity, and spatial extent (measured in units of one spin-flip distance). 

(a) The potential-energy-shell concept allows us to explain the peculiar fact that sample-to-sample fluctuations of imbalance relaxations in small samples, $L\lesssim 20$, are very large at moderate disorder~\cite{EversPRB23}: they are understood as a manifestation of fluctuations of the mass of the shell.

(b) The potential-energy concept implies shell fragmentation above a
finite-size scale $\Wfrag(L)$. This provides a geometric mechanism for the
pronounced slowing down of relaxation and the strong sample-to-sample
fluctuations observed in the disorder regime associated with MBL studies.

(c) The cross-over scale 
$\Wfrag(L)$ shifts with increasing $L$ to larger values, which implies the existence of a parameter region that exhibits a turn-around flow driven by the increasing system size. 
The physical picture behind the flow turning around is the increasing growth (`mass') of the fragment that contains the initial state. In other words, the number of many-body resonances proliferates with increasing $L$, providing more and more relaxation channels.

(d)  We provide a caveat: while fragmentation and many-body localization
are related phenomena, they remain fundamentally different. The full
many-body quantum dynamics also includes tunneling between fragments and
quantum interference. Thus, the dynamics is not confined to a single
fragment. From the perspective of the potential-energy shell, the MBL
problem involves competition between localization on a fragmented backbone
and delocalization due to tunneling, which makes the MBL problem so interesting and challenging.

(e) Since shell fragmentation is a geometric property of the Fock-space graph, its characteristic scale $\Wfrag$ should not be identified with the MBL critical disorder $W_{\mathrm{MBL}}$, which additionally depends on quantum tunneling and interference effects.
It is nevertheless useful to contrast the present Fock-space problem with localization on conventional random regular graphs (RRGs). In the latter, the critical disorder $W_{\mathrm{RRG}}$ depends on a connectivity $K$ that is kept fixed in the thermodynamic limit, and the on-site disorder is uncorrelated~\cite{HerrePRB23}. The Fock-space graph generated by the local many-body Hamiltonian considered here differs in both respects: its connectivity grows with system size, while the Fock-space energies are strongly correlated. These structural differences suggest that the stability and finite-size flow of localization on the present Fock-space graph can differ substantially from those of conventional RRG models~\cite{ThibaultPRB24}. Determining how this distinction ultimately affects the MBL transition requires information beyond that provided by the geometric fragmentation analysis presented here.

\emph{Acknowledgment:}~
Discussions with  I. Gornyi, D. Logan, L. Vidmar, M. Feigelman,  P. Prelovsek, D. Rosa, and D. Sels are gratefully acknowledged. In addition, FE thanks D. Huse, S. Gopalakrishnan, N. O'Dea and R. Bhatt for very useful discussions and feedback during a recent visit.
IM acknowledges financial support from the Prime Minister's Research Fellows (PMRF) scheme offered by the Ministry of Education, Government of India.
SB thanks the National Supercomputing Mission~(NSM) for providing computing resources of `PARAM Porul' at NIT Trichy, and of `PARAM Rudra' at IIT Bombay 
implemented by C-DAC and supported by the Ministry
of Electronics and Information Technology (MeitY) and
Department of Science and Technology~(DST), India.
FE acknowledges that this work was supported by the Deutsche Forschungsgemeinschaft (DFG, German Research Foundation) under EV30/12-1,  
EV30/14-1, 
EV30-16-1, 
further CRC 1277 (Project-ID 314695032, subproject A03), and RTG 2905 (project number 502572516).

\bibliography{MBL}

@article{Cardy99,
title = {Quenched randomness at first-order transitions},
journal = {Physica A: Statistical Mechanics and its Applications},
volume = {263},
number = {1},
pages = {215-221},
year = {1999},
note = {Proceedings of the 20th IUPAP International Conference on Statistical Physics},
issn = {0378-4371},
doi = {https://doi.org/10.1016/S0378-4371(98)00489-0},
url = {https://www.sciencedirect.com/science/article/pii/S0378437198004890},
author = {John Cardy}
}

@article{Cardy96,
   title={Effect of random impurities on fluctuation-driven first-order transitions},
   volume={29},
   ISSN={1361-6447},
   url={http://dx.doi.org/10.1088/0305-4470/29/9/006},
   DOI={10.1088/0305-4470/29/9/006},
   number={9},
   journal={J. Phys. A: Mathematical and General},
   publisher={IOP Publishing},
   author={Cardy, John},
   month=may, 
   note = {1897--1904},
   year={1996}
   }

@article{ZhuWanPRB15,
  title = {Emerging criticality in the disordered three-color Ashkin-Teller model},
  author = {Zhu, Qiong and Wan, Xin and Narayanan, Rajesh and Hoyos, Jos\'e A. and Vojta, Thomas},
  journal = {Phys. Rev. B},
  volume = {91},
  issue = {22},
  pages = {224201},
  numpages = {16},
  year = {2015},
  month = {Jun},
  publisher = {American Physical Society},
  doi = {10.1103/PhysRevB.91.224201},
  url = {https://link.aps.org/doi/10.1103/PhysRevB.91.224201}
}

@article{AbaninAOP21,
title = {{Distinguishing localization from chaos: Challenges in finite-size systems}},
journal = {Ann. Phys.},
volume = {427},
pages = {168415},
year = {2021},
issn = {0003-4916},
doi = {https://doi.org/10.1016/j.aop.2021.168415},
url = {https://www.sciencedirect.com/science/article/pii/S000349162100021X},
author = {D.A. Abanin and J.H. Bardarson and G. {De Tomasi} and S. Gopalakrishnan and V. Khemani and S.A. Parameswaran and F. Pollmann and A.C. Potter and M. Serbyn and R. Vasseur},
}

@ARTICLE{AbaninBloch-Review-2018,
  title = {Colloquium: Many-body localization, thermalization, and entanglement},
  author = {Abanin, Dmitry A. and Altman, Ehud and Bloch, Immanuel and Serbyn, Maksym},
  journal = {Rev. Mod. Phys.},
  volume = {91},
  issue = {2},
  pages = {021001},
  numpages = {26},
  year = {2019},
  month = {May},
  publisher = {American Physical Society},
  doi = {10.1103/RevModPhys.91.021001},
  url = {https://link.aps.org/doi/10.1103/RevModPhys.91.021001}
}

@Article{AletReview2018,
  Title                    = {Many-body localization: An introduction and selected topics},
  Author                   = {F. Alet and N. Laflorencie},
  Journal                  = {C. R. Phys.},
  Year                     = {2018},
  ISSN                     = {1631-0705},
  Url                      = {http://www.sciencedirect.com/science/article/pii/S163107051830032X}
}

@Article{Altshuler1997,
  Title                    = {Quasiparticle Lifetime in a Finite System: A Nonperturbative Approach},
  Author                   = {Altshuler, B.~L. and Gefen, Y. and Kamenev, A. and Levitov, L.~S.},
  Journal                  = {Phys. Rev. Lett.},
  Year                     = {1997},

  Month                    = {Apr},
  Pages                    = {2803--2806},
  Volume                   = {78},

  Doi                      = {10.1103/PhysRevLett.78.2803},
  Issue                    = {14},
  Numpages                 = {0},
  Publisher                = {American Physical Society},
  Url                      = {https://link.aps.org/doi/10.1103/PhysRevLett.78.2803}
}

@article{AltshulerPNAS25,
   title={Renormalization group for Anderson localization on high-dimensional lattices},
   volume={122},
   ISSN={1091-6490},
   url={http://dx.doi.org/10.1073/pnas.2423763122},
   DOI={10.1073/pnas.2423763122},
   number={35},
   journal={Proc. Natl. Acad. Sci. U.S.A.},
   publisher={Proceedings of the National Academy of Sciences},
   author={Altshuler, Boris L. and Kravtsov, Vladimir E. and Scardicchio, Antonello and Sierant, Piotr and Vanoni, Carlo},
   year={2025},
   month=aug }

@Article{Basko2006,
  Title                    = {Metal insulator transition in a weakly interacting many electron system with localized single particle states},
  Author                   = {D.~M. Basko and I.~L. Aleiner and B.~L. Altshuler},
  Journal                  = {Ann. Phys.},
  Year                     = {2006},
  Number                   = {5},
  Pages                    = {1126 - 1205},
  Volume                   = {321},

  ISSN                     = {0003-4916},
  Url                      = {http://www.sciencedirect.com/science/article/pii/S0003491605002630}
}

@Article{Bera2017,
  Title                    = {Density Propagator for Many-Body Localization: Finite-Size Effects, Transient Subdiffusion, and Exponential Decay},
  Author                   = {Bera, S. and De Tomasi, G. and Weiner, F. and Evers, F.},
  Journal                  = {Phys. Rev. Lett.},
  Year                     = {2017},

  Month                    = {May},
  Pages                    = {196801},
  Volume                   = {118},

  Doi                      = {10.1103/PhysRevLett.118.196801},
  Issue                    = {19},
  Numpages                 = {6},
  Publisher                = {American Physical Society},
  Url                      = {https://link.aps.org/doi/10.1103/PhysRevLett.118.196801}
}

@article{BeraRRG18,
  title = {{Return probability for the Anderson model on the random regular graph}},
  author = {Bera, Soumya and De Tomasi, Giuseppe and Khaymovich, Ivan M. and Scardicchio, Antonello},
  journal = {Phys. Rev. B},
  volume = {98},
  issue = {13},
  pages = {134205},
  numpages = {9},
  year = {2018},
  month = {Oct},
  publisher = {American Physical Society},
  doi = {10.1103/PhysRevB.98.134205},
  url = {https://link.aps.org/doi/10.1103/PhysRevB.98.134205}
}

@article{BiroliPRB22,
  title = {{Critical behavior of the Anderson model on the Bethe lattice via a large-deviation approach}},
  author = {Biroli, Giulio and Hartmann, Alexander K. and Tarzia, Marco},
  journal = {Phys. Rev. B},
  volume = {105},
  issue = {9},
  pages = {094202},
  numpages = {11},
  year = {2022},
  month = {Mar},
  publisher = {American Physical Society},
  doi = {10.1103/PhysRevB.105.094202},
  url = {https://link.aps.org/doi/10.1103/PhysRevB.105.094202}
}

@article{BiroliPRB24,
  title = {Large-deviation analysis of rare resonances for the many-body localization transition},
  author = {Biroli, Giulio and Hartmann, Alexander K. and Tarzia, Marco},
  journal = {Phys. Rev. B},
  volume = {110},
  issue = {1},
  pages = {014205},
  numpages = {33},
  year = {2024},
  month = {Jul},
  publisher = {American Physical Society},
  doi = {10.1103/PhysRevB.110.014205},
  url = {https://link.aps.org/doi/10.1103/PhysRevB.110.014205}
}

@article{ColboisPRB24,
  title = {{Statistics of systemwide correlations in the random-field XXZ chain: Importance of rare events in the many-body localized phase}},
  author = {Colbois, Jeanne and Alet, Fabien and Laflorencie, Nicolas},
  journal = {Phys. Rev. B},
  volume = {110},
  issue = {21},
  pages = {214210},
  numpages = {25},
  year = {2024},
  month = {Dec},
  publisher = {American Physical Society},
  doi = {10.1103/PhysRevB.110.214210},
  url = {https://link.aps.org/doi/10.1103/PhysRevB.110.214210}
}

@article{CreedPRB23,
  title = {{Probability transport on the Fock space of a disordered quantum spin chain}},
  author = {Creed, Isabel and Logan, David E. and Roy, Sthitadhi},
  journal = {Phys. Rev. B},
  volume = {107},
  issue = {9},
  pages = {094206},
  numpages = {19},
  year = {2023},
  month = {Mar},
  publisher = {American Physical Society},
  doi = {10.1103/PhysRevB.107.094206},
  url = {https://link.aps.org/doi/10.1103/PhysRevB.107.094206}
}

@article{detomasiRRGPRB20,
  title = {{Subdiffusion in the Anderson model on the random regular graph}},
  author = {De Tomasi, Giuseppe and Bera, Soumya and Scardicchio, Antonello and Khaymovich, Ivan M.},
  journal = {Phys. Rev. B},
  volume = {101},
  issue = {10},
  pages = {100201},
  numpages = {7},
  year = {2020},
  month = {Mar},
  publisher = {American Physical Society},
  doi = {10.1103/PhysRevB.101.100201},
  url = {https://link.aps.org/doi/10.1103/PhysRevB.101.100201}
}

@article{detomasiVinftyPRB19,
  title = {{Dynamics of strongly interacting systems: From Fock-space fragmentation to many-body localization}},
  author = {De Tomasi, Giuseppe and Hetterich, Daniel and Sala, Pablo and Pollmann, Frank},
  journal = {Phys. Rev. B},
  volume = {100},
  issue = {21},
  pages = {214313},
  numpages = {15},
  year = {2019},
  month = {Dec},
  publisher = {American Physical Society},
  doi = {10.1103/PhysRevB.100.214313},
  url = {https://link.aps.org/doi/10.1103/PhysRevB.100.214313}
}

@article{detomasiFSPRB21,
  title = {{Rare thermal bubbles at the many-body localization transition from the Fock space point of view}},
  author = {De Tomasi, Giuseppe and Khaymovich, Ivan M. and Pollmann, Frank and Warzel, Simone},
  journal = {Phys. Rev. B},
  volume = {104},
  issue = {2},
  pages = {024202},
  numpages = {13},
  year = {2021},
  month = {Jul},
  publisher = {American Physical Society},
  doi = {10.1103/PhysRevB.104.024202},
  url = {https://link.aps.org/doi/10.1103/PhysRevB.104.024202}
}

@article{LongPRL23,
  title = {Phenomenology of the Prethermal Many-Body Localized Regime},
  author = {Long, David M. and Crowley, Philip J. D. and Khemani, Vedika and Chandran, Anushya},
  journal = {Phys. Rev. Lett.},
  volume = {131},
  issue = {10},
  pages = {106301},
  numpages = {7},
  year = {2023},
  month = {Sep},
  publisher = {American Physical Society},
  doi = {10.1103/PhysRevLett.131.106301},
  url = {https://link.aps.org/doi/10.1103/PhysRevLett.131.106301}
}

@article{ChavezUltraSlow23,
  title = {Ultraslow Growth of Number Entropy in an $\ensuremath{\ell}$-Bit Model of Many-Body Localization},
  author = {Aceituno Ch\'avez, David and Artiaco, Claudia and Klein Kvorning, Thomas and Herviou, Lo\"{\i}c and Bardarson, Jens H.},
  journal = {Phys. Rev. Lett.},
  volume = {133},
  issue = {12},
  pages = {126502},
  numpages = {7},
  year = {2024},
  month = {Sep},
  doi = {10.1103/PhysRevLett.133.126502}
}

@Article{Doggen2018,
  title = {Many-body localization and delocalization in large quantum chains},
  author = {Doggen, Elmer V. H. and Schindler, Frank and Tikhonov, Konstantin S. and Mirlin, Alexander D. and Neupert, Titus and Polyakov, Dmitry G. and Gornyi, Igor V.},
  journal = {Phys. Rev. B},
  volume = {98},
  issue = {17},
  pages = {174202},
  numpages = {13},
  year = {2018},
  month = {Nov},
  publisher = {American Physical Society},
  doi = {10.1103/PhysRevB.98.174202},
  url = {https://link.aps.org/doi/10.1103/PhysRevB.98.174202}
}

@article{EversPRB23,
  title = {Internal clock of many-body delocalization},
  author = {Evers, Ferdinand and Modak, Ishita and Bera, Soumya},
  journal = {Phys. Rev. B},
  volume = {108},
  issue = {13},
  pages = {134204},
  numpages = {15},
  year = {2023},
  month = {Oct},
  publisher = {American Physical Society},
  doi = {10.1103/PhysRevB.108.134204},
  url = {https://link.aps.org/doi/10.1103/PhysRevB.108.134204}
}

@article{GarciaPRL17,
  title = {{Scaling Theory of the Anderson Transition in Random Graphs: Ergodicity and Universality}},
  author = {Garc\'{\i}a-Mata, I. and Giraud, O. and Georgeot, B. and Martin, J. and Dubertrand, R. and Lemari\'e, G.},
  journal = {Phys. Rev. Lett.},
  volume = {118},
  issue = {16},
  pages = {166801},
  numpages = {5},
  year = {2017},
  month = {Apr},
  publisher = {American Physical Society},
  doi = {10.1103/PhysRevLett.118.166801},
  url = {https://link.aps.org/doi/10.1103/PhysRevLett.118.166801}
}

@article{GarciaMataRRGPRR20,
  title = {{Two critical localization lengths in the Anderson transition on random graphs}},
  author = {Garc\'{\i}a-Mata, I. and Martin, J. and Dubertrand, R. and Giraud, O. and Georgeot, B. and Lemari\'e, G.},
  journal = {Phys. Rev. Res.},
  volume = {2},
  issue = {1},
  pages = {012020},
  numpages = {7},
  year = {2020},
  month = {Jan},
  publisher = {American Physical Society},
  doi = {10.1103/PhysRevResearch.2.012020},
  url = {https://link.aps.org/doi/10.1103/PhysRevResearch.2.012020}
}

@Article{Gornyi2005,
  Title                    = {{Interacting Electrons in Disordered Wires: Anderson Localization and Low-$T$ Transport}},
  Author                   = {Gornyi, I. V. and Mirlin, A. D. and Polyakov, D. G.},
  Journal                  = {Phys. Rev. Lett.},
  Year                     = {2005},

  Month                    = {Nov},
  Pages                    = {206603},
  Volume                   = {95},

  Doi                      = {10.1103/PhysRevLett.95.206603},
  Issue                    = {20},
  Numpages                 = {4},
  Publisher                = {American Physical Society},
  Url                      = {https://link.aps.org/doi/10.1103/PhysRevLett.95.206603}
}

@article{HerrePRB23,
  title = {Ergodicity-to-localization transition on random regular graphs with large connectivity and in many-body quantum dots},
  author = {Herre, Jan-Niklas and Karcher, Jonas F. and Tikhonov, Konstantin S. and Mirlin, Alexander D.},
  journal = {Phys. Rev. B},
  volume = {108},
  issue = {1},
  pages = {014203},
  numpages = {21},
  year = {2023},
  month = {Jul},
  publisher = {American Physical Society},
  url = {https://link.aps.org/doi/10.1103/PhysRevB.108.014203}
}

@Article{Khemani2017,
  Title                    = {Critical Properties of the Many-Body Localization Transition},
  Author                   = {Khemani, V. and Lim, S.~P. and Sheng, D.~N. and Huse, D.~A.},
  Journal                  = {Phys. Rev. X},
  Year                     = {2017},

  Month                    = {Apr},
  Pages                    = {021013},
  Volume                   = {7},

  Doi                      = {10.1103/PhysRevX.7.021013},
  Issue                    = {2},
  Numpages                 = {11},
  Publisher                = {American Physical Society},
  Url                      = {https://link.aps.org/doi/10.1103/PhysRevX.7.021013}
}

@article{MaceMultifractality2018,
  title = {Multifractal Scalings Across the Many-Body Localization Transition},
  author = {Mac\'e, Nicolas and Alet, Fabien and Laflorencie, Nicolas},
  journal = {Phys. Rev. Lett.},
  volume = {123},
  issue = {18},
  pages = {180601},
  numpages = {6},
  year = {2019},
  month = {Oct},
  publisher = {American Physical Society},
  doi = {10.1103/PhysRevLett.123.180601},
  url = {https://link.aps.org/doi/10.1103/PhysRevLett.123.180601}
}

@article{Morningstar2022,
  title = {Avalanches and many-body resonances in many-body localized systems},
  author = {Morningstar, Alan and Colmenarez, Luis and Khemani, Vedika and Luitz, David J. and Huse, David A.},
  journal = {Phys. Rev. B},
  volume = {105},
  issue = {17},
  pages = {174205},
  numpages = {20},
  year = {2022},
  month = {May},
  publisher = {American Physical Society},
  doi = {10.1103/PhysRevB.105.174205},
  url = {https://link.aps.org/doi/10.1103/PhysRevB.105.174205}
}

@misc{maceira2024,
      title={Thermalization Dynamics in Closed Quantum Many Body Systems: a Precision Large Scale Exact Diagonalization Study}, 
      author={Ivo A. Maceira and Andreas M. L\"auchli},
      year={2024},
      eprint={2409.18863},
      archivePrefix={arXiv},
      primaryClass={quant-ph},
      url={https://arxiv.org/abs/2409.18863}, 
}

@article{NandyPRB21,
  title = {Dephasing in strongly disordered interacting quantum wires},
  author = {Nandy, Sourav and Evers, Ferdinand and Bera, Soumya},
  journal = {Phys. Rev. B},
  volume = {103},
  issue = {8},
  pages = {085105},
  numpages = {11},
  year = {2021},
  month = {Feb},
  publisher = {American Physical Society},
  doi = {10.1103/PhysRevB.103.085105},
  url = {https://link.aps.org/doi/10.1103/PhysRevB.103.085105}
}

@article{pietracaprina2019hilbert,
title = {Hilbert-space fragmentation, multifractality, and many-body localization},
journal = {Ann. Phys.},
volume = {435},
pages = {168502},
year = {2021},
note = {Special Issue on Localisation 2020},
issn = {0003-4916},
doi = {https://doi.org/10.1016/j.aop.2021.168502},
url = {https://www.sciencedirect.com/science/article/pii/S0003491621001081},
author = {Francesca Pietracaprina and Nicolas Laflorencie}
}

@ARTICLE{PandaMBL19,
	doi = {10.1209/0295-5075/128/67003},
	url = {https://doi.org/10.1209%2F0295-5075%2F128%2F67003},
	year = 2020,
	month = {feb},
	publisher = {{IOP} Publishing},
	volume = {128},
	number = {6},
	pages = {67003},
	author = {R. K. Panda and A. Scardicchio and M. Schulz and S. R. Taylor and M. {\v{Z}}nidari{\v{c}}},
	title = {Can we study the many-body localisation transition?},
	journal = {{EPL} (Europhysics Letters)}
}

@article{PrasadPRB22,
  title = {Initial state dependent dynamics across the many-body localization transition},
  author = {Prasad, Yogeshwar and Garg, Arti},
  journal = {Phys. Rev. B},
  volume = {105},
  issue = {21},
  pages = {214202},
  numpages = {13},
  year = {2022},
  month = {Jun},
  publisher = {American Physical Society},
  doi = {10.1103/PhysRevB.105.214202},
  url = {https://link.aps.org/doi/10.1103/PhysRevB.105.214202}
}

@article{PrelovsekPRB16,
  title = {Dynamical conductivity and its fluctuations along the crossover to many-body localization},
  author = {Bari\ifmmode \check{s}\else \v{s}\fi{}i\ifmmode \acute{c}\else \'{c}\fi{}, Osor S. and Kokalj, Jure and Balog, Ivan and Prelov\ifmmode \check{s}\else \v{s}\fi{}ek, Peter},
  journal = {Phys. Rev. B},
  volume = {94},
  issue = {4},
  pages = {045126},
  numpages = {6},
  year = {2016},
  month = {Jul},
  publisher = {American Physical Society},
  doi = {10.1103/PhysRevB.94.045126},
  url = {https://link.aps.org/doi/10.1103/PhysRevB.94.045126}
}

@article{PrelovsekPercPRB21,
  title = {Many-body localization as a percolation phenomenon},
  author = {Prelov\ifmmode \check{s}\else \v{s}\fi{}ek, P. and Mierzejewski, M. and Krsnik, J. and Bari\ifmmode \check{s}\else \v{s}\fi{}i\ifmmode \acute{c}\else \'{c}\fi{}, O. S.},
  journal = {Phys. Rev. B},
  volume = {103},
  issue = {4},
  pages = {045139},
  numpages = {8},
  year = {2021},
  month = {Jan},
  publisher = {American Physical Society},
  doi = {10.1103/PhysRevB.103.045139},
  url = {https://link.aps.org/doi/10.1103/PhysRevB.103.045139}
}

@Article{Polkovnikov2021,
  Title                    = {Dynamical obstruction to localization in a disordered spin chain},
  Author                   = {Sels, D. and Polkovnikov, A.},
  Journal                  = {Phys. Rev. E},
  Year                     = {2021},

  Month                    = {Nov},
  Pages                    = {054105},
  Volume                   = {104},
  

  Doi                      = {https://doi.org/10.1103/PhysRevE.104.054105},
  Numpages                 = {11},
  Publisher                = {American Physical Society},
  Url                      = {https://journals.aps.org/pre/abstract/10.1103/PhysRevE.104.054105}
}

@misc{rizzoBethe24,
      title={{The localized phase of the Anderson model on the Bethe lattice}}, 
      author={Tommaso Rizzo and Marco Tarzia},
      year={2024},
      eprint={2406.18748},
      archivePrefix={arXiv},
      primaryClass={cond-mat.dis-nn},
      url={https://arxiv.org/abs/2406.18748}, 
}

@article{RoyReview_2025,
doi = {10.1088/1361-648X/ad94c3},
url = {https://dx.doi.org/10.1088/1361-648X/ad94c3},
year = {2024},
month = {dec},
publisher = {IOP Publishing},
volume = {37},
number = {7},
pages = {073003},
author = {Roy, Sthitadhi and Logan, David E},
title = {{The Fock-space landscape of many-body localisation}},
journal = {J. Phys. Condens. Matter}
}

@article{RoyPercPRB19,
  title = {{Percolation in Fock space as a proxy for many-body localization}},
  author = {Roy, Sthitadhi and Chalker, J. T. and Logan, David E.},
  journal = {Phys. Rev. B},
  volume = {99},
  issue = {10},
  pages = {104206},
  numpages = {16},
  year = {2019},
  month = {Mar},
  publisher = {American Physical Society},
  doi = {10.1103/PhysRevB.99.104206},
  url = {https://link.aps.org/doi/10.1103/PhysRevB.99.104206}
}

@article{RoyPercPRB19a,
  title = {Exact solution of a percolation analog for the many-body localization transition},
  author = {Roy, Sthitadhi and Logan, David E. and Chalker, J. T.},
  journal = {Phys. Rev. B},
  volume = {99},
  issue = {22},
  pages = {220201},
  numpages = {5},
  year = {2019},
  month = {Jun},
  publisher = {American Physical Society},
  doi = {10.1103/PhysRevB.99.220201},
  url = {https://link.aps.org/doi/10.1103/PhysRevB.99.220201}
}

@article{NieddaRG2024,
  title = {Renormalization group analysis of the many-body localization transition in the random-field XXZ chain},
  author = {Niedda, Jacopo and Testasecca, Giacomo Bracci and Magnifico, Giuseppe and Balducci, Federico and Vanoni, Carlo and Scardicchio, Antonello},
  journal = {Phys. Rev. B},
  volume = {112},
  issue = {14},
  pages = {144201},
  numpages = {14},
  year = {2025},
  month = {Oct},
  publisher = {American Physical Society},
  doi = {10.1103/gcwf-jdlr},
  url = {https://link.aps.org/doi/10.1103/gcwf-jdlr}
}

@Article{SierantSci23,
	title={{Universality in Anderson localization on random graphs with varying connectivity}},
	author={Piotr Sierant and Maciej Lewenstein and Antonello Scardicchio},
	journal={SciPost Phys.},
	volume={15},
	pages={045},
	year={2023},
	publisher={SciPost},
	doi={10.21468/SciPostPhys.15.2.045},
	url={https://scipost.org/10.21468/SciPostPhys.15.2.045},
}

@article{ScoquartPRB25,
  title = {{Scaling of many-body localization transitions: Quantum dynamics in Fock space and real space}},
  author = {Scoquart, Thibault and Gornyi, Igor V. and Mirlin, Alexander D.},
  journal = {Phys. Rev. B},
  volume = {112},
  issue = {6},
  pages = {064203},
  numpages = {32},
  year = {2025},
  month = {Aug},
  publisher = {American Physical Society},
  doi = {10.1103/qcvd-n2yk},
  url = {https://link.aps.org/doi/10.1103/qcvd-n2yk}
}

@Article{Schreiber2015,
  Title                    = {Observation of many-body localization of interacting fermions in a quasirandom optical lattice},
  Author                   = {Schreiber, M. and Hodgman, S.~S. and Bordia, P. and L{\"u}schen, H.~P. and Fischer, M.~H. and Vosk, R. and Altman, E. and Schneider, U. and Bloch, I.},
  Journal                  = {Science},
  Year                     = {2015},
  Number                   = {6250},
  Pages                    = {842--845},
  Volume                   = {349},
  Doi                      = {10.1126/science.aaa7432},
  ISSN                     = {0036-8075},
  Publisher                = {American Association for the Advancement of Science},
  Url                      = {http://science.sciencemag.org/content/349/6250/842}
}

@article{SierantReview_2025,
doi = {10.1088/1361-6633/ad9756},
url = {https://dx.doi.org/10.1088/1361-6633/ad9756},
year = {2025},
month = {jan},
publisher = {IOP Publishing},
volume = {88},
number = {2},
pages = {026502},
author = {Sierant, Piotr and Lewenstein, Maciej and Scardicchio, Antonello and Vidmar, Lev and Zakrzewski, Jakub},
title = {{Many-body localization in the age of classical computing}},
journal = {{Rep. Prog. Phys.}}
}

@article{SierantPRB22,
  title = {Challenges to observation of many-body localization},
  author = {Sierant, Piotr and Zakrzewski, Jakub},
  journal = {Phys. Rev. B},
  volume = {105},
  issue = {22},
  pages = {224203},
  numpages = {18},
  year = {2022},
  month = {Jun},
  publisher = {American Physical Society},
  doi = {10.1103/PhysRevB.105.224203},
  url = {https://link.aps.org/doi/10.1103/PhysRevB.105.224203}
}

@article{SirkerPRL20,
  title = {Evidence for Unbounded Growth of the Number Entropy in Many-Body Localized Phases},
  author = {Kiefer-Emmanouilidis, Maximilian and Unanyan, Razmik and Fleischhauer, Michael and Sirker, Jesko},
  journal = {Phys. Rev. Lett.},
  volume = {124},
  issue = {24},
  pages = {243601},
  numpages = {6},
  year = {2020},
  month = {Jun},
  publisher = {American Physical Society},
  doi = {10.1103/PhysRevLett.124.243601},
  url = {https://link.aps.org/doi/10.1103/PhysRevLett.124.243601}
}

@article{SonnerPRB17,
  title = {Multifractality of wave functions on a Cayley tree: From root to leaves},
  author = {Sonner, M. and Tikhonov, K. S. and Mirlin, A. D.},
  journal = {Phys. Rev. B},
  volume = {96},
  issue = {21},
  pages = {214204},
  numpages = {15},
  year = {2017},
  month = {Dec},
  publisher = {American Physical Society},
  doi = {10.1103/PhysRevB.96.214204},
  url = {https://link.aps.org/doi/10.1103/PhysRevB.96.214204}
}

@article{SuntasPRB20,
  title = {Ergodicity breaking transition in finite disordered spin chains},
  author = {\v{S}untajs, J. and Bon\v{c}a, J. and Prosen, T. and Vidmar, L.},
  journal = {Phys. Rev. B},
  volume = {102},
  issue = {6},
  pages = {064207},
  numpages = {11},
  year = {2020},
  month = {Aug},
  publisher = {American Physical Society},
  doi = {10.1103/PhysRevB.102.064207},
  url = {https://link.aps.org/doi/10.1103/PhysRevB.102.064207}
}

@article{SutradharPRB22,
  title = {{Scaling of the Fock-space propagator and multifractality across the many-body localization transition}},
  author = {Sutradhar, Jagannath and Ghosh, Soumi and Roy, Sthitadhi and Logan, David E. and Mukerjee, Subroto and Banerjee, Sumilan},
  journal = {Phys. Rev. B},
  volume = {106},
  issue = {5},
  pages = {054203},
  numpages = {10},
  year = {2022},
  month = {Aug},
  publisher = {American Physical Society},
  doi = {10.1103/PhysRevB.106.054203},
  url = {https://link.aps.org/doi/10.1103/PhysRevB.106.054203}
}

@article{TikhonovBethePRB16,
  title = {Fractality of wave functions on a Cayley tree: Difference between tree and locally treelike graph without boundary},
  author = {Tikhonov, K. S. and Mirlin, A. D.},
  journal = {Phys. Rev. B},
  volume = {94},
  issue = {18},
  pages = {184203},
  numpages = {16},
  year = {2016},
  month = {Nov},
  publisher = {American Physical Society},
  doi = {10.1103/PhysRevB.94.184203},
  url = {https://link.aps.org/doi/10.1103/PhysRevB.94.184203}
}

@article{TikhonovAnn21,
title = {{From Anderson localization on random regular graphs to many-body localization}},
journal = {Ann. Phys.},
volume = {435},
pages = {168525},
year = {2021},
note = {Special Issue on Localisation 2020},
issn = {0003-4916},
doi = {https://doi.org/10.1016/j.aop.2021.168525},
url = {https://www.sciencedirect.com/science/article/pii/S0003491621001317},
author = {K.S. Tikhonov and A.D. Mirlin}
}

@article{TikhonovRRG16,
  title = {Anderson localization and ergodicity on random regular graphs},
  author = {Tikhonov, K. S. and Mirlin, A. D. and Skvortsov, M. A.},
  journal = {Phys. Rev. B},
  volume = {94},
  issue = {22},
  pages = {220203},
  numpages = {6},
  year = {2016},
  month = {Dec},
  publisher = {American Physical Society},
  doi = {10.1103/PhysRevB.94.220203},
  url = {https://link.aps.org/doi/10.1103/PhysRevB.94.220203}
}

@article{ThibaultPRB24,
  title = {{Role of Fock-space correlations in many-body localization}},
  author = {Scoquart, Thibault and Gornyi, Igor V. and Mirlin, Alexander D.},
  journal = {Phys. Rev. B},
  volume = {109},
  issue = {21},
  pages = {214203},
  numpages = {26},
  year = {2024},
  month = {Jun},
  publisher = {American Physical Society},
  doi = {10.1103/PhysRevB.109.214203},
  url = {https://link.aps.org/doi/10.1103/PhysRevB.109.214203}
}

@article{Torres_SelfPRB20,
  title = {Self-averaging in many-body quantum systems out of equilibrium: Approach to the localized phase},
  author = {Torres-Herrera, E. Jonathan and De Tomasi, Giuseppe and Schiulaz, Mauro and P\'erez-Bernal, Francisco and Santos, Lea F.},
  journal = {Phys. Rev. B},
  volume = {102},
  issue = {9},
  pages = {094310},
  numpages = {13},
  year = {2020},
  month = {Sep},
  publisher = {American Physical Society},
  doi = {10.1103/PhysRevB.102.094310},
  url = {https://link.aps.org/doi/10.1103/PhysRevB.102.094310}
}

@article{VanoniRRG24,
   title={{Renormalization group analysis of the Anderson model on random regular graphs}},
   volume={121},
   ISSN={1091-6490},
   url={http://dx.doi.org/10.1073/pnas.2401955121},
   DOI={10.1073/pnas.2401955121},
   number={29},
   journal={Proc. Natl. Acad. Sci. U.S.A.},
   publisher={Proceedings of the National Academy of Sciences},
   author={Vanoni, Carlo and Altshuler, Boris L. and Kravtsov, Vladimir E. and Scardicchio, Antonello},
   year={2024},
   month=jul }

@article{Weiner19,
  title = {Slow dynamics and strong finite-size effects in many-body localization with random and quasiperiodic potentials},
  author = {Weiner, Felix and Evers, Ferdinand and Bera, Soumya},
  journal = {Phys. Rev. B},
  volume = {100},
  issue = {10},
  pages = {104204},
  numpages = {11},
  year = {2019},
  month = {Sep},
  publisher = {American Physical Society},
  doi = {10.1103/PhysRevB.100.104204},
  url = {https://link.aps.org/doi/10.1103/PhysRevB.100.104204}
}

@Misc{SuppMat,
  Note                  = {See the Supplemental Material for the dynamical correlation in the fragmented regime, the noninteracting $V=0$ data, the fragmented landscape for other initial states, and a discussion of fragmentation in a power-law interacting model, which additionally includes Refs.~\cite{MirlinPRBMPRB00, maceira2024, RoyPercPRB19, PrelovsekPercPRB21, TikhonovRRG16, VanoniRRG24, AltshulerPNAS25, detomasiVinftyPRB19, BurinPRB15, Mirlin_longrange2018, EversReview08, MirlinPRBMPRB00} in addition to other reference already mentioned above.}
}

@article{SunPRB25,
  title = {{Characterizing dynamical criticality of many-body localization transitions from a Fock-space perspective}},
  author = {Sun, Zheng-Hang and Wang, Yong-Yi and Cui, Jian and Fan, Heng and Heyl, Markus},
  journal = {Phys. Rev. B},
  volume = {111},
  issue = {9},
  pages = {094210},
  numpages = {12},
  year = {2025},
  month = {Mar},
  publisher = {American Physical Society},
  doi = {10.1103/PhysRevB.111.094210},
  url = {https://link.aps.org/doi/10.1103/PhysRevB.111.094210}
}

@article{BurinPRB15,
  title = {Many-body delocalization in a strongly disordered system with long-range interactions: Finite-size scaling},
  author = {Burin, Alexander L.},
  journal = {Phys. Rev. B},
  volume = {91},
  issue = {9},
  pages = {094202},
  numpages = {8},
  year = {2015},
  month = {Mar},
  publisher = {American Physical Society},
  doi = {10.1103/PhysRevB.91.094202},
  url = {https://link.aps.org/doi/10.1103/PhysRevB.91.094202}
}

@article{Mirlin_longrange2018,
  title = {Many-body localization transition with power-law interactions: Statistics of eigenstates},
  author = {Tikhonov, K. S. and Mirlin, A. D.},
  journal = {Phys. Rev. B},
  volume = {97},
  issue = {21},
  pages = {214205},
  numpages = {11},
  year = {2018},
  month = {Jun},
  publisher = {American Physical Society},
  doi = {10.1103/PhysRevB.97.214205},
  url = {https://link.aps.org/doi/10.1103/PhysRevB.97.214205}
}

@article{EversReview08,
  title = {Anderson transitions},
  author = {Evers, Ferdinand and Mirlin, Alexander D.},
  journal = {Rev. Mod. Phys.},
  volume = {80},
  issue = {4},
  pages = {1355--1417},
  numpages = {0},
  year = {2008},
  month = {Oct},
  publisher = {American Physical Society},
  doi = {10.1103/RevModPhys.80.1355},
  url = {https://link.aps.org/doi/10.1103/RevModPhys.80.1355}
}

@article{MirlinPRBMPRB00,
  title = {Multifractality and critical fluctuations at the Anderson transition},
  author = {Mirlin, A. D. and Evers, F.},
  journal = {Phys. Rev. B},
  volume = {62},
  issue = {12},
  pages = {7920--7933},
  numpages = {0},
  year = {2000},
  month = {Sep},
  publisher = {American Physical Society},
  doi = {10.1103/PhysRevB.62.7920},
  url = {https://link.aps.org/doi/10.1103/PhysRevB.62.7920}
}


\clearpage

\appendix

\makeatletter
\renewcommand{\appendixname}{}
\renewcommand{\thesection}{{S\arabic{section}}}
\renewcommand{\thefigure}{S\arabic{figure}}
\renewcommand{\theequation}{\thesection.\arabic{equation}}
 
\setcounter{page}{1}
\setcounter{figure}{0}
\makeatother
\maketitle

\onecolumngrid

\centerline{\bf Supplemental Material}
\medskip
\centerline{for}
\medskip
\centerline{\large{\bf Fock space fragmentation in quenches of disordered interacting fermions}}
\medskip
\centerline{by Ishita Modak, Rajesh Narayanan, Ferdinand Evers, and Soumya Bera}
\bigskip
\bigskip
\twocolumngrid
\tableofcontents
\bigskip
\bigskip

\section{Discussion -- Relation to literature} 
\label{Appendix_literature}
(i) It has recently been reported that in the absence of disorder, the energy of the initial state and its energy variance play a crucial role in determining thermalization in closed clean quantum systems~\cite {maceira2024}.
As one would expect, the energy variance sets the timescale over which local observables approach their thermal values for a clean non-integrable model. 
The parameters $E_\text{in}$ and $\Delta_\text{in}$ will also play a similar role in the disordered case.

(ii) Roy, Chalker, and Logan have introduced an alternative Fock space connectivity concept to describe MBL as a percolation problem in Fock space~\cite{RoyPercPRB19a, RoyPercPRB19}.
They consider a Fock space Hamiltonian
\begin{align}
    \hat{\msr{H}} \coloneqq \sum_{b} E_b |b\rangle\langle b| + \sum_{b\neq b'} T_{b b^\prime} |b\rangle\langle b'|  \text{.}
\end{align}
Two ``nodes" $|b\rangle,|b^\prime\rangle$ are considered as linked (an edge is ``active") if the difference of two many-body energies falls below the matrix element for hybridization (resonance condition): 
\begin{align} 
|E_b-E_{b^\prime}|\lesssim |T_{bb^\prime}|. 
\label{e6} 
\end{align} 
Two nodes are in the same percolation cluster if they are connected by a continuous sequence of active links. The percolation-cluster concept differs from the potential-energy shell: The condition \eqref{e6} defines a local bound, i.e., for each node, there is another node nearby in energy. This implies that, taking an arbitrary pair of nodes out of the percolation cluster, the resonance condition \eqref{e6} will not necessarily be satisfied. 
In contrast, the condition eq. 3 from the main text acts as a global bound: any pair of sites from the potential-energy shell is close in energy in the sense of eq. 3 from the main text. 
Because of this difference, it remains to be seen to what extent the statistical properties of potential energy shells and percolation clusters, which share the same average energy per node, resemble each other. 

(iii) In a related work \textcite{pietracaprina2019hilbert} investigated the MBL transition by introducing a decimation method motivated by the multifractal nature of the many-body eigenstate in the Fock-space. It leads to fragmentation as in the previous study~\cite{RoyPercPRB19}, however, here the authors introduced a cutoff parameter, which controls the fragmentation in Eq.~\eqref{e6}; i.e., the vertices that are kept depend on the parameter $\Lambda$
$
|E_b-E_b^\prime|/|T_{b b^\prime}| < \Lambda
$. 
Our work again differed here due to the very nature of the protocol and the additional correlation between fragmentation and transient dynamics that we uncover~(see \ref{cls_dyn}).

(iv) \textcite{PrelovsekPercPRB21} also analyzed the MBL problem from the Fock-space percolation perspective. The basis, however, is not the occupation basis; rather, it is constructed from the non-interacting Anderson orbitals, i.e., $\hat{T}+\hat{U}$
is diagonal~(refer Eq.1 from he main text). On that basis, interaction $\hat{V}$ plays the role of hopping in the high-dimensional graph. Analyzing the resonant cluster structure on this basis, the authors observed percolation-like phenomena, where clusters of sites decoupled from the rest in the large disorder regime, $W\gtrsim 6$. Consequently, the transient dynamics is slow, which we also observe even at weak disorder $W \approx 1.5$. 
Finally, the authors concluded that the disappearance of a single macroscopic cluster signals the MBL transition; in contrast, we observe a crossover from a weakly fragmented `energy shell' to a regime where the fragmentation of the energy shell is extreme for a specific initial state.

(v) The Fock-space graph shown in Fig.~1  invites comparison with random regular graphs (RRGs), which have been widely used as effective models of high-dimensional localization.
The statistical properties of wavefunctions on RRGs have attracted considerable attention over recent years~\cite{TikhonovRRG16, GarciaPRl17, BeraRRG18, GarciaMataRRGPRR20, TikhonovAnn21, SierantSci23, VanoniRRG24, AltshulerPNAS25}. 
Note that the RRG-ensemble and the associated scaling properties analyzed by us differ significantly from the traditional literature studies. 
First, we investigate a genuine Fock-space dynamics generated by local Hamiltonians, so Fock-space disorder is exponentially correlated in the following sense. While the number of eigenvalues $E_b$ of $\hat H^\prime$ is exponentially large, $\sim e^{\alpha L}$ with $\alpha$ of order unity, these eigenvalues cannot be mutually independent because the Hamiltonian (eq. 1 from the main text) features only $L$ uncorrelated parameters $\epsilon_x, x=1,\ldots,L$.

Second, following the physical process, we take a thermodynamic limit in which the connectivity of the graph, $K$, is not invariant but scales with the logarithm of the graph size, i.e., the Fock-space volume: $K\approx L\sim \ln(\text{\Ndim})$. 
We consider both of these points as qualitatively relevant for understanding the charge-dynamics given with the model Eq.1 from the main text.

(vi) We also mention that the idea of Fock-space fragmentation for t-V model (refer eq. 1 from the main text) has been discussed previously by \textcite{detomasiVinftyPRB19}. The fragmentation is observed in the limit $\thop/V\rightarrow 0$. In that work, the Fock-space fragments into several invariant subspaces due to the kinetic constraints, which is different than what we discuss here in the context of quench, where the role of the initial state is paramount for fragmentation to observe in the limit $\thop/V=1$.

\begin{figure}[!tb]
\includegraphics[width=1.0\columnwidth]{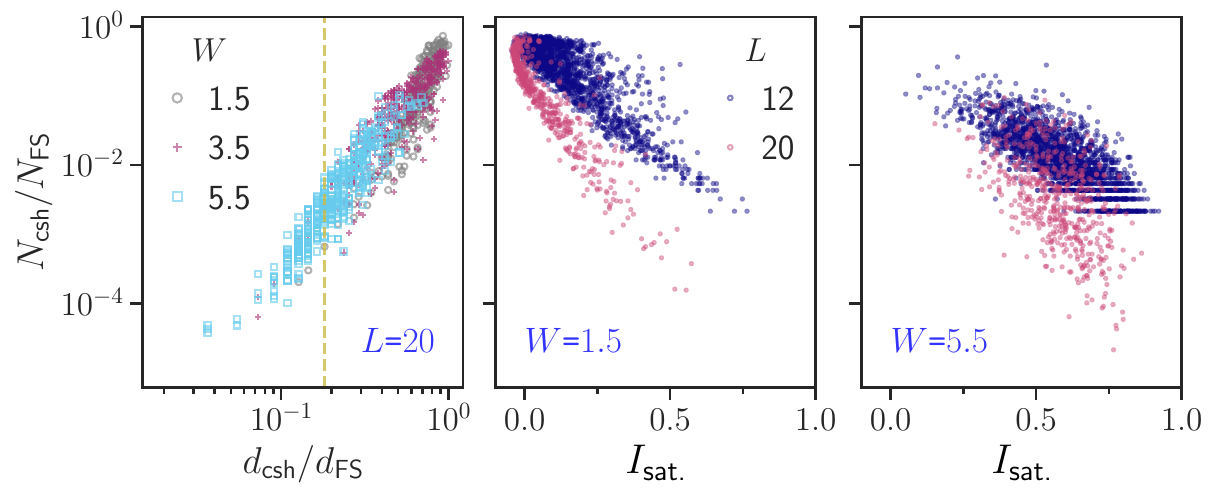}
    \caption{Correlations between the normalized mass of the cluster, \Nc, which embeds the initial state, and the long-time system dynamics. Each data point represents one out of ${\sim} 3 {\times} 10^3$  samples. 
    Left: \Nc over the size of the embedding cluster measured in terms of the single spin flip distance. Especially at large disorder, the correlation between mass and size of the embedding cluster is seen to be weak. 
    Center/Right: Correlation between \Nc and the steady-state imbalance, $I_\text{sat.}= I(t\to\infty)$, for two system sizes $L\Equal12, 20$, and disorder values 
    $W\Equal 1.5, 5.5 $. 
    \label{f7}}
\end{figure}


\section{Cluster geometry and long-time dynamics} 
\label{cls_dyn}
In order to shed light into the relation between the geometry of the cluster, which is embedding the initial state, and the system dynamics, we have correlated the mass of this cluster, \Nc, with its extension throughout the Fock space in Fig. \ref{f7}, left; the extension is given in terms of the maximum spin-flip distance away from the initial state reached in this cluster.  
This distance, $d_\text{csh}$,  is of interest here, because one expects thermalization if this distance is comparable to the total extent of the Fock-space:
 $d_\text{csh}\approx d_\text{FS}$. 

The correlation between \Nc and $d_\text{csh}$ is seen to be pronounced - as one might have expected: inspecting Fig. \ref{f7}, left, a rough  phenomenological parametrization can be devised, e.g., along the lines of 
$\overline{\text{\Nc/\Ndim}}{\sim} 
[(d_\text{csh})/d_\text{FS}]^\delta, 
\delta\approx 3.6$.
The slow variation of the cluster size with \Nc manifests as a small exponent $1/\delta$; due to the fluctuations of \Nc seen in Fig. \ref{f7}, left, a given size $d_\text{csh}$ can be realized with connected clusters varying in mass \Nc by an order of magnitude.

Hence, dynamical indicators, - e.g. the saturation value of the imbalance $I_\text{sat.}\coloneqq I(t{\to}\infty)$ -  correlate relatively weakly with \Nc, especially at strong disorder, see Fig. \ref{f7}, right, but better with the normalized distance $d_\text{csh}/d_\text{max}$, see Fig.~\ref{f7a}. The fluctuations are large for $W=3.5$, as previously observed.

\begin{figure}[!tb]
    \centering
    \includegraphics[width=1.0\columnwidth]{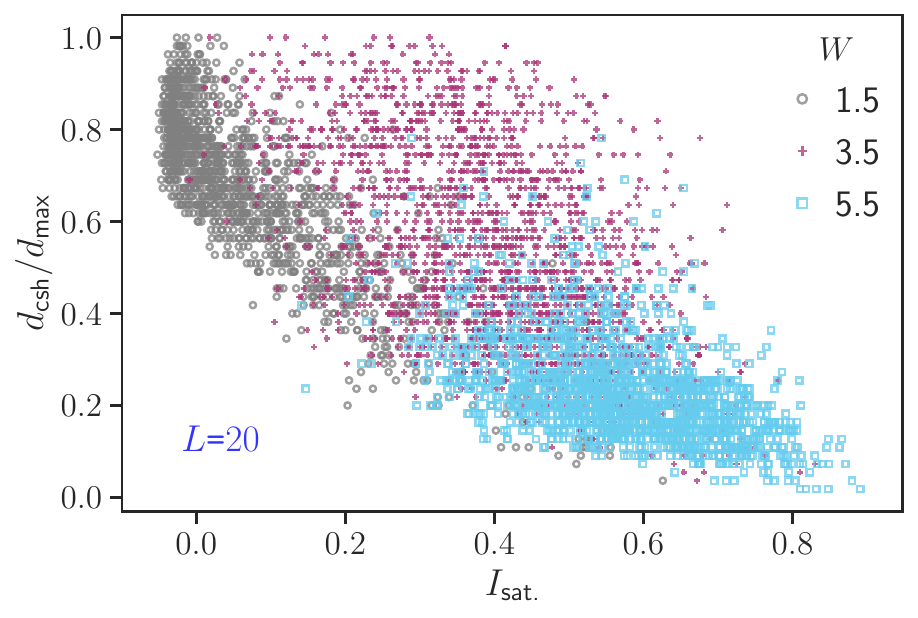}   
    \caption{Correlation between normalized distance $d_\text{csh}/d_\text{max}$ of the connected cluster with imabalnce saturation $I_\text{sat.}$ for different $W$. The most scattering in the data is visible at $W=3.5$.}
    \label{f7a}
\end{figure}

In passing, we mention that correlation patterns, such as seen in Fig. \ref{f7}, exhibit a pronounced flow with system size, which is a manifestation of the evolution discussed before with Fig. 5 in the main text. 
Two snap-shots for $L=12,20$ are presented in Fig. \ref{f7}, central for intermediate disorder $W{=}1.5$. As compared to the dark ``cloud" ($L=12$), the light cloud ($L=20$) has migrated towards the upper left, indicating a contraction to the weak-coupling fixed point \Nc/\Ndim$=1$ at $I_\text{sat.}=0$. 
%

\section{Fragmentation and dynamical observables} 
\label{cls_dyn}

\begin{figure*}[t]
    \centering
    \includegraphics[width=1.\textwidth]{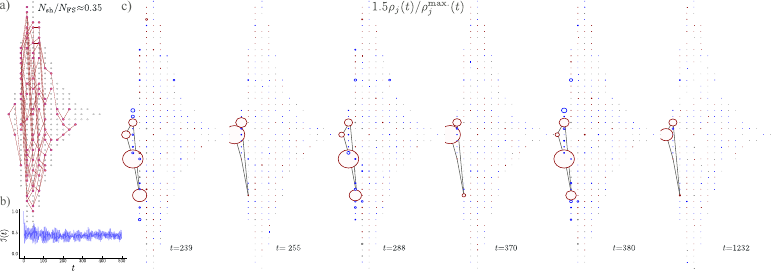}
    \caption{Distribution of FS sites within the energy shell for a representative disorder realization at $W \Equal 1.5$ for $L=10$, showing pronounced fragmentation. (b)~The corresponding time evolution of the imbalance $I(t)$ reflects this structural feature, remaining large and exhibiting oscillations. (c)~Wavefunction amplitudes $\rho_j(t)$, normalized by their maximum value at each time, are shown across FS sites. From intermediate times ($t \sim 10^2$) to long times ($t \sim 10^3$), $\rho_j(t)$ displays persistent oscillatory behavior confined to the fragmented region. These recurring patterns are the dynamical resonances mentioned in the text. }
    \label{fig:resonance}
\end{figure*}

%

\subsection{Transient dynamics: resonances and echos}
The Fock-space perspective Fig. 3 of the main text allows for the convenient identification of phenomena that are manifestations of fragmentation in time-dependent observables, especially on transient time scales.
Fig. ~\ref{fig:resonance} shows the wavefunction amplitudes $\rho_j(t) \Equal |\Psi_j(t)|^2$ on FS sites $j$ at various times for a given disorder realization, along with the corresponding FS connectivity graph and the time evolution of the imbalance $I(t)$ in the inset. 
This sample contains approximately 35\% sites inside the energy shell, \Nred, in FS but exhibits a fragmented structure with only four sites that include the N\'eel state: \Nc$=4$.  
Since tunneling is weak, even at long times ($t \gtrsim 10^3$) the wavefunction remains confined within this small cluster, which in this sense defines the ``dynamically active space". 

Since \Nc is small, the spatial pattern of $\rho_j(t)$ in FS shows strong recurrences -  a signature of resonances within the fragment.
Such resonances manifest as pronounced oscillations - ``echos" - in the imbalance as illustrated in the insets of Fig. \ref{fig:resonance} or Fig. 3 of the main text.

\subsection{Further discussion on fragmentation}
(a) Fragmentation of the potential-energy shell appears to be the defining characteristic of the strong-disorder regime. While it implies a dramatic slow-down of relaxation processes, it does not necessarily imply many-body localization in view of residual tunneling processes.  

(b) We consider the ratio of variance and mean, 
\begin{align}
    \mathfrak{r}_\text{csh}\coloneqq \frac{\text{var}(\text{\Nc}/\text{\Nred})}{\overline{\text{\Nc}/\text{\Nred}}} = \sqrt{L} \frac{f(\overline{\text{\Nc}/\text{\Nred}})}{\overline{\text{\Nc}/\text{\Nred}}}, 
\end{align}
which indicates self-averaging if 
$\mathfrak{r}_\text{csh}\to 0$ in the limit of large $L$. From Fig. 5 of the main text, we extract that indeed $\mathfrak{r}_\text{csh}$ vanishes at the weak-disorder fixed point, while otherwise $\mathfrak{r}_\text{csh}\propto \sqrt{L}$, in particular, also at strong disorder, i.e., in the regime of fragmentation. 
In other words, fluctuations of the size of the cluster that embeds the initial state are so large in the fragmented regime that an average cluster size is not indicative of the typical situation.

(c) Irrespective of whether or not the strong-disorder fixpoint seen in Fig. 
5 of the main text, is unstable; there is a wide window of intermediate system sizes, in which fragmentation is a relevant concept to understand the transient relaxation dynamics. This regime features intermediate time scales, $\tau_\text{c}$, that reflect the quantum dynamics exploring the fragment. They manifest as strong fluctuations such as echoes in time-series; an example is shown in the lower inset of Fig. 3 of the main text. Following the fluctuations in the cluster size, \Nc, also the fluctuations in $\tau_\text{c}$ are expected to be large.

(d) The fragments are connected amongst each other via residual tunnel couplings, i.e., hopping processes that invoke sites (denoted as grey in Fig. 3 of the mian text) outside the potential-energy shell. These couplings provide a timescale, $\tau_\text{creep}$, beyond which the dynamics begins to explore the neighborhood of the fragment. Following our previous work\cite{Weiner19,NandyPRB21} on the relaxation of quenches at very long times, we identify $\tau_\text{creep}$ with the onset of creep.

\begin{figure}[!tb]
    \centering
    \includegraphics[width=1.0\columnwidth]{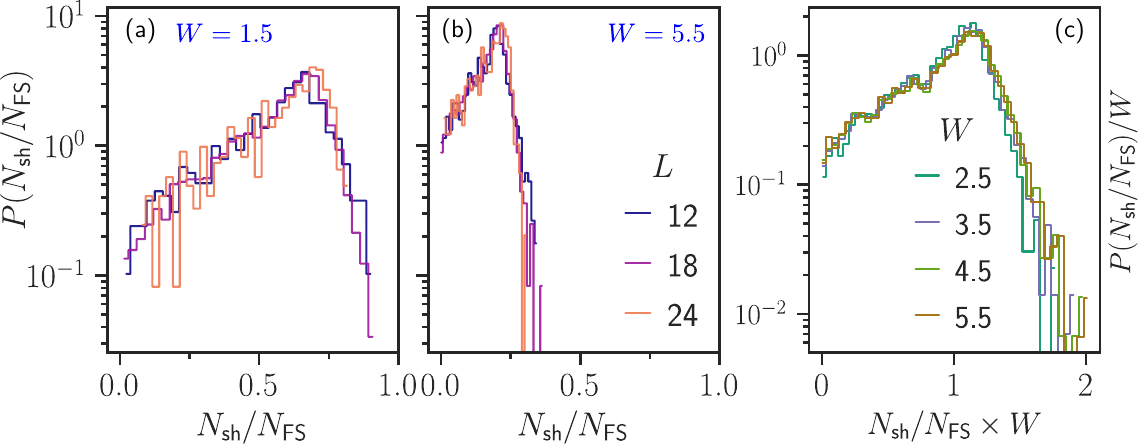}
    \caption{(a-b) Data similar to Fig. 2 of the main text for the non-interacting case, $V=0$ for different $W$, and $L$. (c) Shows an approximate collapse of the full distribution with rescaled x, and y axes with $W$. This distribution is independent of $L$, and here is shown for $L=18$.}
    \label{fig:nshell_ndim_dist_ni}
\end{figure}
\section{Non-interacting analysis}
\subsection{Distribution of $N_{\mathrm{sh}}/N_{\mathrm{dim}}$ for $V=0.0$}

The distribution of \Nred/\Ndim is independent of the system size $L$, as seen in Fig.~\ref{fig:nshell_ndim_dist_ni} (a-b) for $V=0$. The maximum of the distribution shifts towards \Nred/\Ndim $\rightarrow 0$ with increasing $W$, signaling a decrease in cluster mass. 

Figure~\ref{fig:nshell_ndim_dist_ni}(c) shows an approximate collapse to a universal distribution with rescaled x- and y-axis with $W$. The movement of the peak as observed in the left panels is thus $\propto W$. Within the Gaussian model introduced in the main text, this is seen from the scaling of the width of the initial energy $\Delta_\text{in} = (4/3 (W/4)^2 L)^{1/2} $; the fluctuation is dominated by $W$ in the limit $W/\thop \gtrsim 1$. 

The finite-size effects that gave rise to strong sample-to-sample fluctuations at weak disorder in the interacting model $V=1$~(see Fig. 2 of the main text) are absent here. In the regime of $W$, where $\xi \ll L$, where $\xi$ is the non-interacting localization length, the effect of interaction is already relatively small as observed in the main text. 
\begin{figure}[!tbh]
    \centering
    \includegraphics[width=0.8\columnwidth]{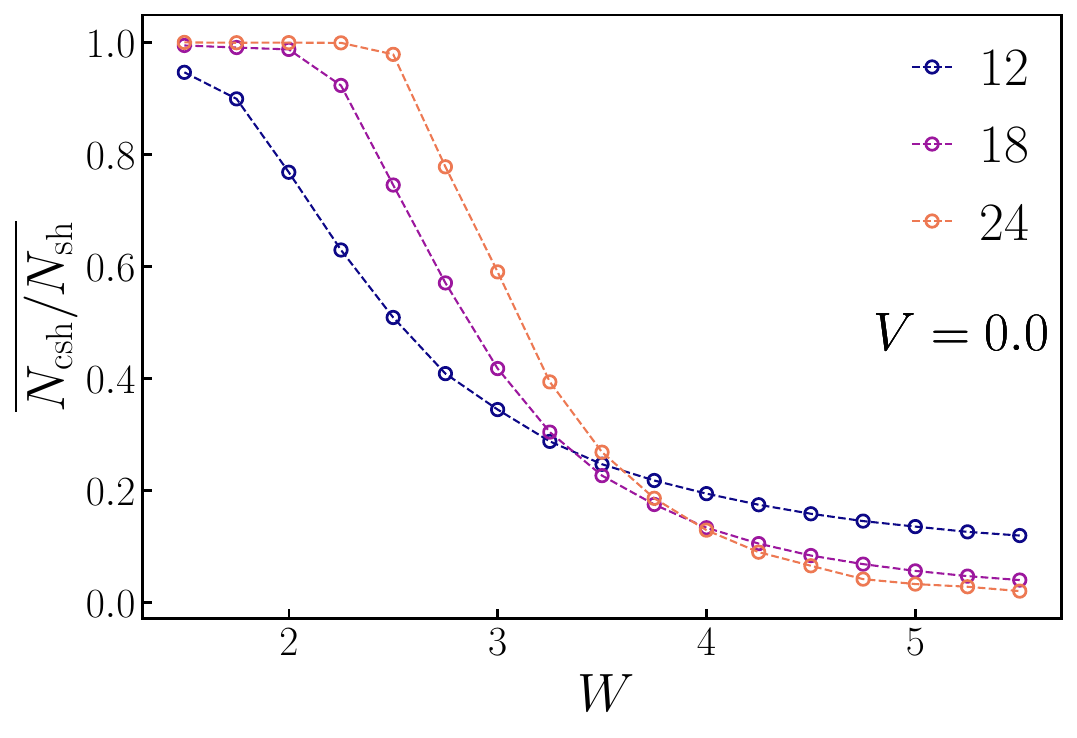}
    \caption{Data similar to Fig. 4 of the main text for the non-interacting case, $V=0$. 
    }
    \label{fig:NI_fragment}
\end{figure}

\begin{figure}[!tbh]
    \centering
\includegraphics[width=0.8\columnwidth]{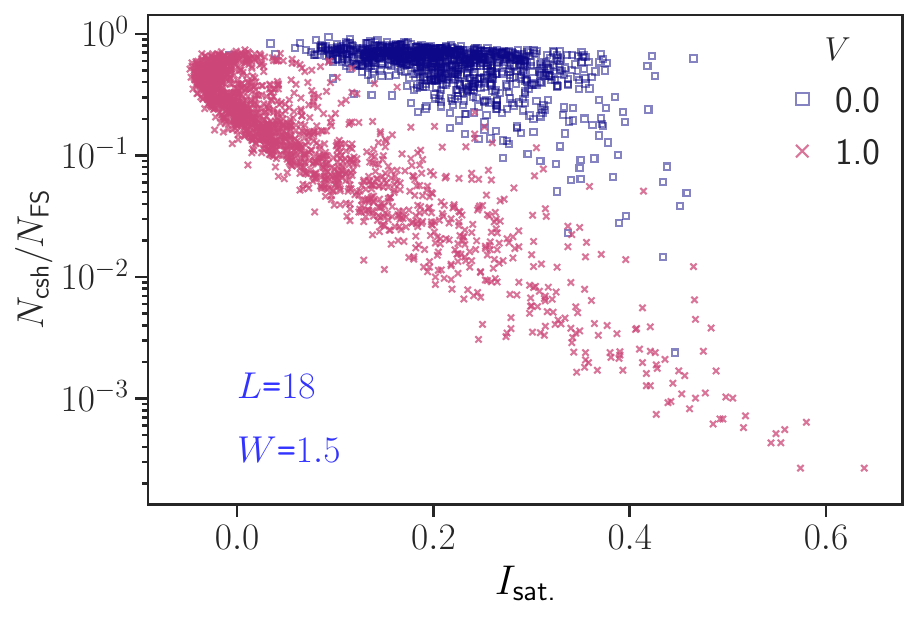}
    \caption{Comparison between interacting and non-interacting  correlation of \Nc/\Ndim with the long time imbalance saturation value \Isat for $W=1.5$.}
    \label{fig:isat_ncsh_corr}
\end{figure}
\subsection{Fragmentation in the non-interacting limit} 
Figure~\ref{fig:NI_fragment} shows the \Nc/\Nred dependence with $W$ for $V=0$. The data exhibit a similar crossover behavior, as observed for the interacting model (Fig. 4 from the main text), which is not surprising given that this regime is dominated by $W$. More precisely, Fig.~\ref{fig:NI_fragment} emphasizes that the crossover to fragmentation in Fock-space is almost dominated by non-interacting physics exactly in the regime where the non-interacting length scale is the lattice spacing, $\xi/a \sim 1$.

Nonetheless, we make the following observation that \Nc/\Nred  being unity does not imply thermalization either for finite $V$ or $V=0$. The thermalization should be checked independently, as it is shown in Fig. A1 from the main text appendix. 
The role of interaction is to make the system thermal in the weak coupling regime \Nc/\Nred$\sim 1$.

\subsection{Long time dynamics in the non-interacting limit}
We compare the correlation between the \Nc/\Ndim with the long-time saturation value of imbalance, i.e. \Isat in Fig. \ref{fig:isat_ncsh_corr}. 
We observe a clustering of data points belonging to $V=0$ around $I_{\mathrm{sat}}\approx 0.15$. However, the corresponding $V=1$ data points exhibit a large fluctuation in \Isat values, with a substantially large number of data points clustering around \Isat$\approx 0$, indicating thermalization. 

The weak correlation between \Nred/\Ndim with \Isat that we observed in the strong disorder $W\gtrsim 3$ in Fig. A1 from the main text appendix, now can be understood by reconciling Figs.~\ref{fig:NI_fragment} and \ref{fig:isat_ncsh_corr} - the effect of interaction at such disorder strength is too weak to observe the significant flow of the data in the left upper corner (i.e., $I_{\mathrm{sat}}\rightarrow  0., \, \text{\Nc/\Ndim} \rightarrow 1$ as seen in Fig.~\ref{fig:isat_ncsh_corr} for $W=1.5$), which implies thermalization.
%

\section{A remark on self-averaging} 

Early work on MBL has performed a DMRG study of the random-field $S=1/2$ Heisenberg model and established self-averaging of the observable sublattice imbalance at large enough system sizes, typically $L\gtrsim 50$ spins~\cite{Doggen2018}. This study has addressed observation times up to $t=100$. 
However, recent ED-study invoking much longer observation times up to $t\simeq 10^4$ has challenged the self-averaging property at moderate disorder~\cite{Torres_SelfPRB20}: the authors concluded a violation of self-averaging for the imbalance $I(t)$ in the long-time limit. 

We clarify the relation between the two papers in the following way. 
The spreading of the saturation values $I_\text{sat}$ displayed in Fig. A1, from the main text, is much broader than the average $\overline{I_\text{sat}}$. Hence, indeed, the imbalance is not self-averaging in the range of system sizes shown in Fig. A1, in the main text. Therefore, we confirm the conclusion of \textcite{Torres_SelfPRB20} that self-averaging is much more difficult to achieve in the long-time limit. However, since \textcite{Torres_SelfPRB20}  have been working at system sizes up to $L=16$, they effectively considered the limit of $t\to\infty$ first at fixed $L$ in their study. 
It is the inverted limit of $L\to\infty$ first, which is the one relevant for the traditional concept of self-averaging. This case was investigated in Ref.~\cite{Doggen2018} and here self-averaging is restored - also in the weak-disorder (``chaotic")  regime. Our results, Fig. 5 of the main text, agree with this conclusion. 
\section{Fock space landscapes}
\subsection{$\alpha$-dependence}
\begin{figure}[!bh]
    \centering
\includegraphics[width=1.0\columnwidth]{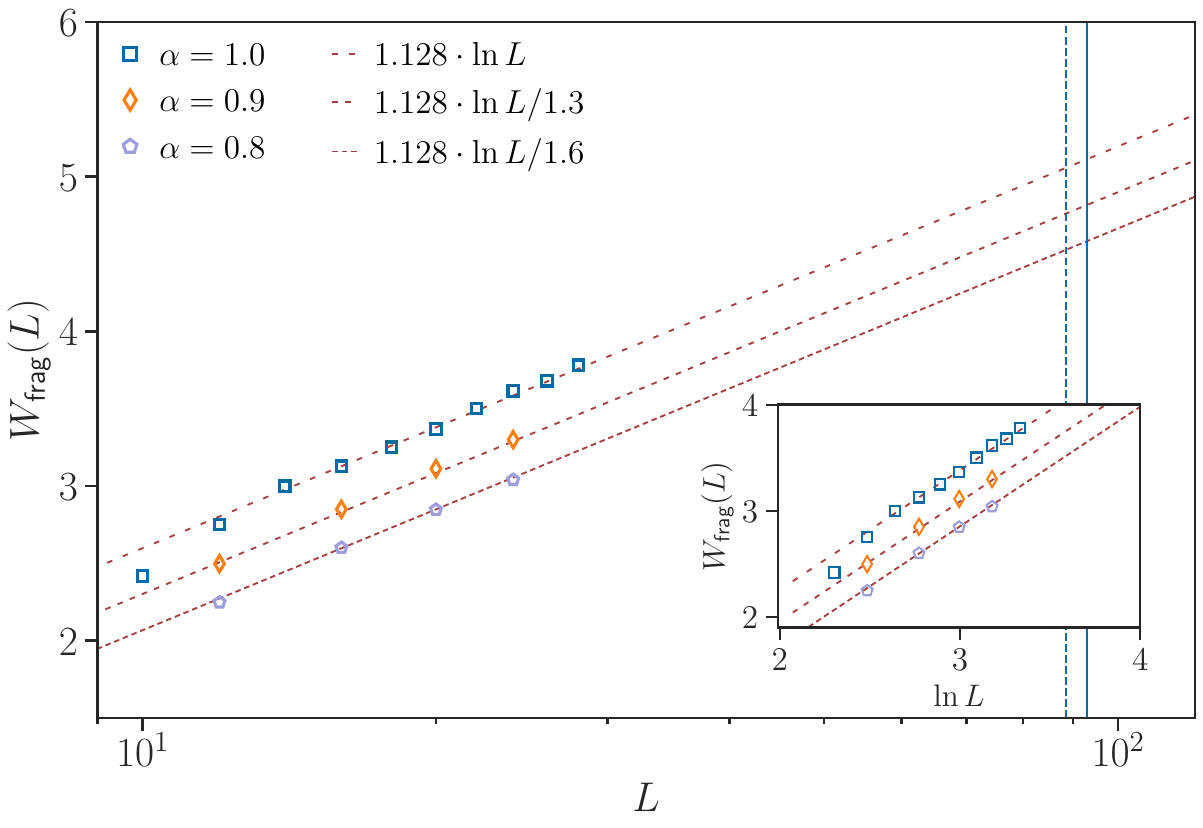}
    \caption{Shows the difference between different choices of $\alpha$, which determines the width of the energy shell. The inset shows a zoomed view of the data, highlighting the logarithmic flow of $\Wfrag(L)$ with $L$. The critical $W$ changes from $\sim 5.2$ for $\alpha = 1$ to $\sim 4.5$ for $\alpha = 0.8$. The vertical line denotes $L_*(\alpha = 1) \simeq 93 $, while the dashed vertical line denotes $L_*(\alpha = 0.8) \simeq  88$.}
    \label{fig:alpha_d}
\end{figure}
In the main text, the width of the energy shell is defined as, 
\begin{align}
    |E_\text{in}-E_b|\lesssim \alpha \Delta_\text{in}, 
\end{align}
where, $\Delta_\text{in} \coloneqq (\braPsiin \hat{\msr{H}}^2-E_\text{in}^2\ketPsiin)^{1/2}$ and $\alpha$ a phenomenological parameter. Here we examine the dependence of $\Wfrag(L)$ on $\alpha$, where $\Wfrag(L)$ defines the finite size fragmentation length scale as characterized by the maximum fluctuation of $\text{\Nc/\Nred}$.

As seen in Fig.~\ref{fig:alpha_d}, the logarithmic flow of $\Wfrag(L)$ with $L$ persists for all values of $\alpha$. 
Following the same procedure, changing the shell parameter from $\alpha=1$ to $\alpha=0.8$, corresponding
to a $20\%$ change, shifts the finite-size estimate of $W_c$ from $5.2$ to
$4.5$ (about $13.5\%$), while the corresponding estimate of the turnaround
length changes only from $L_*=93$ to $88$ (about $5.4\%$).

This suggests that our estimate for $\Wfrag$ in the asymptotic limit is an upper bound for the fragmentation transition of the energy shell.

\subsection{The state $|1100\ldots\rangle$ }
To test the generality of the Fock-space energy-shell landscape, we analyzed another initial state, $|110011\ldots\rangle$. We observe similar sample-to-sample fluctuations in the Fock-space energy-shell size, initial state energy, and their corresponding effects on imbalance dynamics, as shown in Fig.~\ref{fig:dup_state}. 

\begin{figure*}[!tbh]
    \centering
    \includegraphics[width=0.9\textwidth]{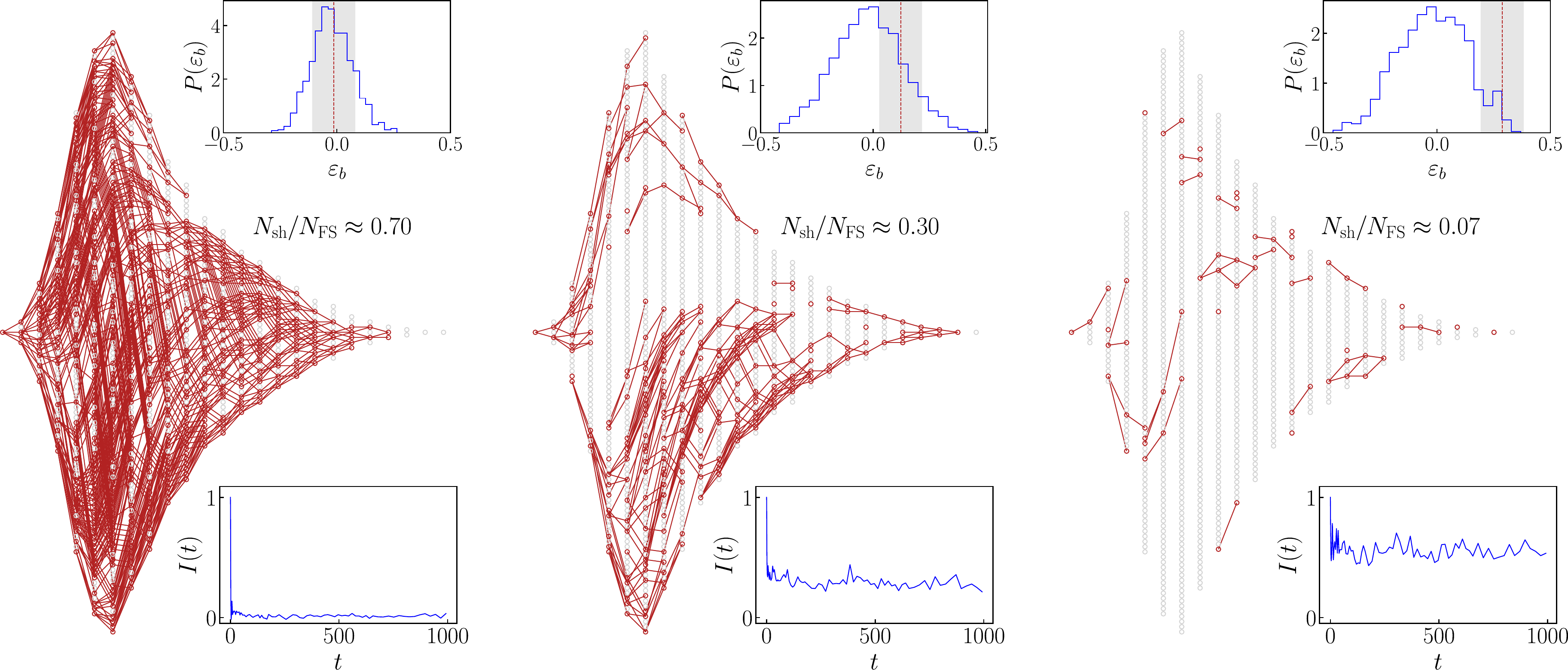}
    \caption{The Fock-space fragmentation structure for three different samples for initial state $|110011\ldots\rangle$ for $W=1.5$ for $L=12$. A similar correlation is observed, with increasing fragmentation of the potential energy shell, the imbalance $I(t)$ becomes finite.}
    \label{fig:dup_state}
\end{figure*}


\subsection{Analysis of the initial state $|000 \textcolor{blue}{010101} 111\ra$}
\begin{figure*}[!tb]
    \centering
    \includegraphics[width=0.995\linewidth]{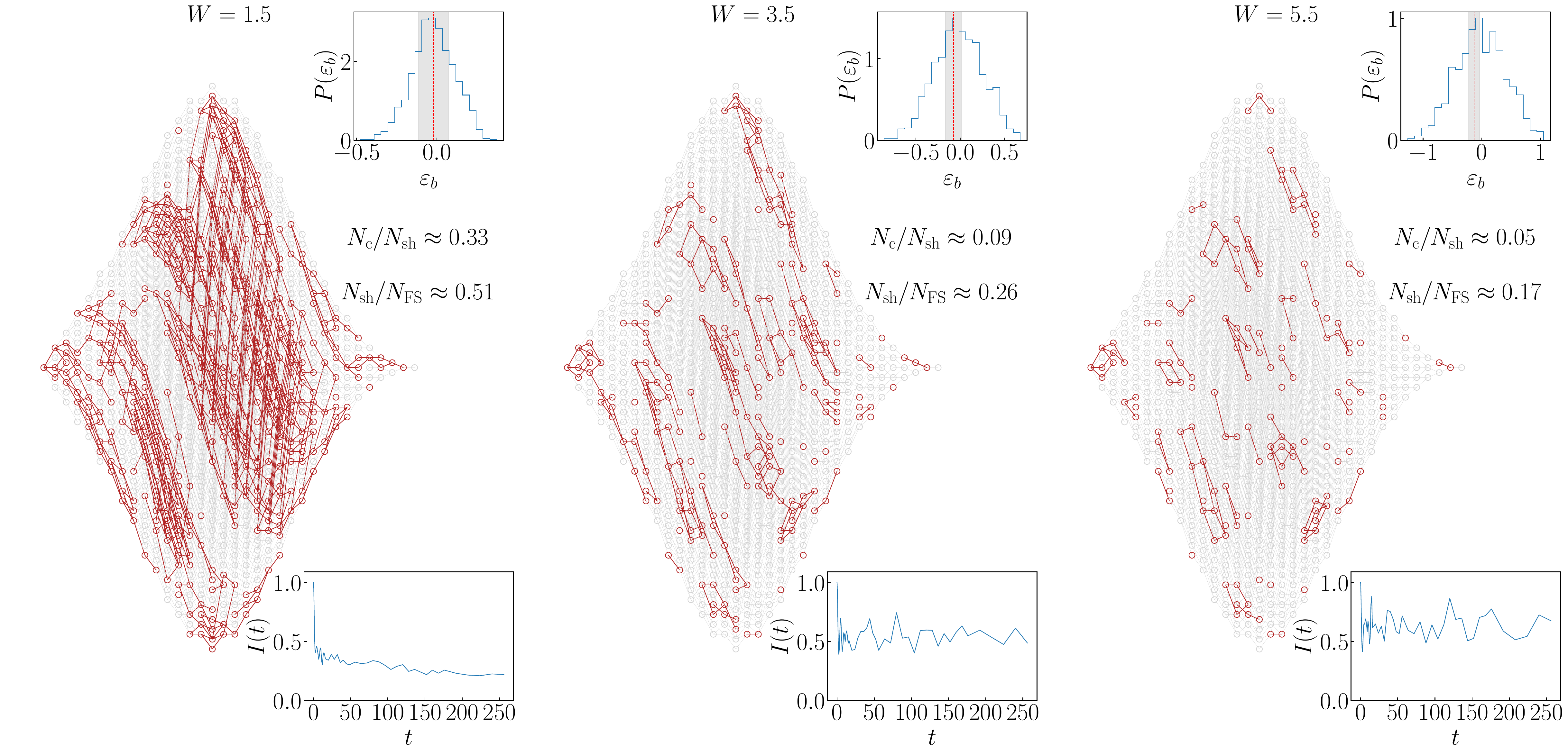}
    \caption{The Fock-space fragmentation structure for a single sample for three different disorder strengths for the initial state $|000 \textcolor{blue}{010101} 111\ra$ for L = 12. Upper inset: The distribution of the Fock space node energies. Lower inset: Dynamics of the particle density imbalance defined as ${I}(t) = \sum_{i=3L/4}^{L} \langle\hat{n}_i(t)\rangle - \sum_{i=1}^{L/4} \langle\hat{n}_i(t)\rangle + \sum_{i\in [L/4+1, 3L/4-2]} (\langle\hat{n}_{i+1}(t)\rangle - \langle\hat{n}_i(t)\rangle)$.
} 
\label{fig:another_state}
\end{figure*}

We have already shown that the initial state $|11001100\ldots\ra$ exhibits the same fragmentation behavior with increasing disorder strength as observed in the main text. We include in Fig.~\ref{fig:another_state} another example for an initial state exhibiting a smaller number of domain walls placed at uncorrelated positions, i.e., $|000 \textcolor{black}{010101} 111\ra$. 

\begin{figure*}[!htb]
    \centering
    \includegraphics[width=1\linewidth]{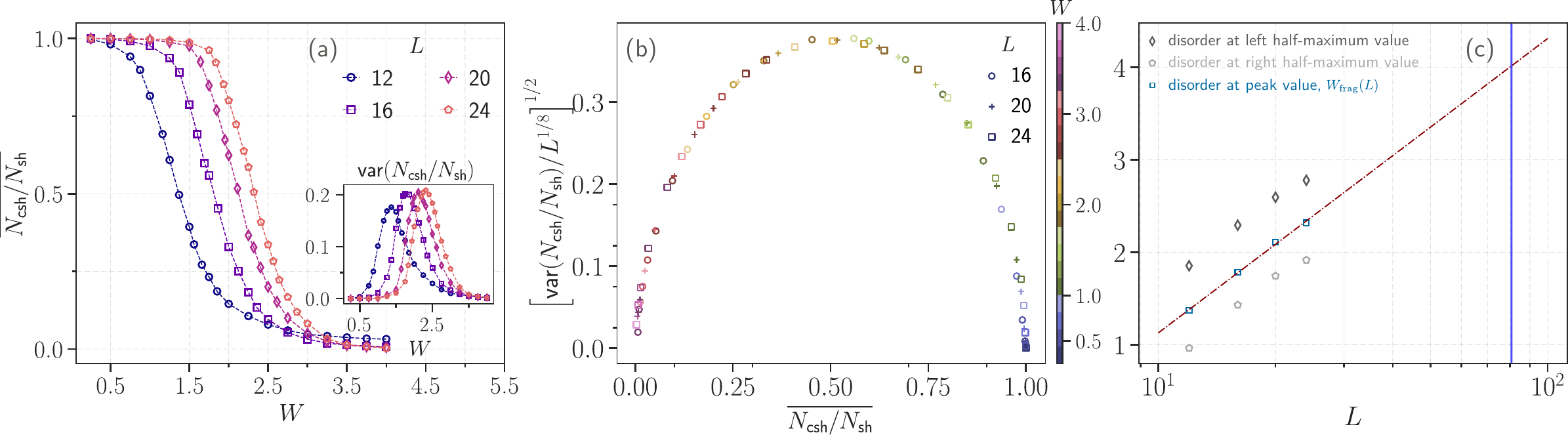}
    \caption{Analysis of the $\Wfrag(L)$ for an initial state with $L/4$ number of domains. (a) Shows the flow of $\text{\Nc/\Nred}$ with $W$. Inset shows the flow of $\text{var}(\text{\Nc/\Nred})$. (b) The collapse of the data that is shown in (a) with the help of a heuristic parameter $L^{1/8}$. (c) Shows the variation of $\Wfrag(L)$ with $L$. }
    \label{fig:mean_var_as}
\end{figure*}

In Fig.~\ref{fig:mean_var_as}, we perform the scaling analysis for an initial state with only $L/4$ domain walls. The data again exhibits a similar flow of $\text{\Nc/\Nred}$ with $L$, and a collapse of $\overline{\text{\Nc/\Nred}}$ against $\text{var}(\text{\Nc/\Nred})$ upon rescaling with a phenomenological factor $L^{1/8}$ (middle panel). This exponent differs from that of the N\'eel state, where the optimal scaling is obtained with $L^{1/2}$.

The logarithmic flow of $\Wfrag(L)$, shown in the right panel, is qualitatively similar to that of the N\'eel state, yielding an estimate of the critical disorder strength $\Wfrag \approx 4$. The discrepancy of the $\Wfrag$ with the N\'eel could be attributed to the different energy content of the initial state.

\section{Fragmentation in a power-law interacting model}
\begin{figure*}[!htb]
    \centering
    \includegraphics[width=0.95\linewidth]{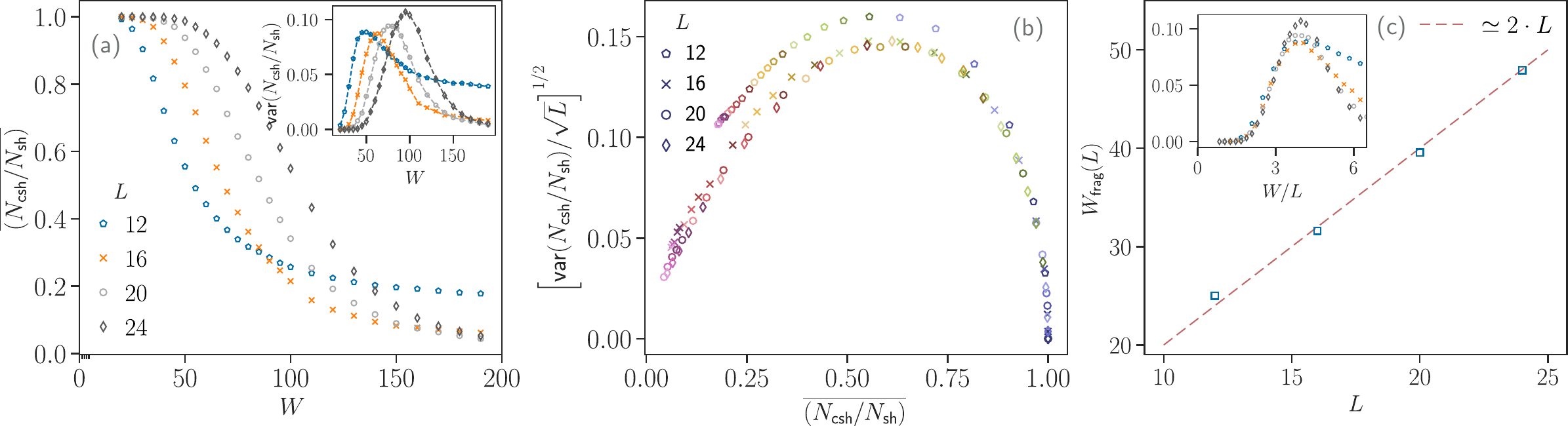}
    \caption{Fragmentation analysis for the long range model with $\delta=3/2$~\eqref{eq:long_range_ham}.
(a) Main panel shows the $\overline{\text{\Nc/\Nred}}$ as a function of $W$ for different system sizes $L$. Inset shows $\mathrm{var}(\text{\Nc/\Nred})$ plotted as a function of the disorder strength $W$. 
(b) Shows the scaling collapse of $\overline{\text{\Nc/\Nred}}$ against $\mathrm{var}(\text{\Nc/\Nred})$, where the variance is rescaled by a heuristic factor $\sqrt{L}$.
(c) Shows the extrapolation of the flow of $\Wfrag(L)$ with $L$. This fit yields a flow of the $\Wfrag(L) \propto L$. Inset shows the collapse of the $\mathrm{var}(\text{\Nc/\Nred})$ data when plotted with rescaled disorder strength $W/L$.}
\label{fig:long_range}
\end{figure*}
We next demonstrate that the qualitative features we report are also present in more general classes of Hamiltonians that exhibit power-law interactions - rather than the nearest-neighbor interactions characteristic of the XXZ Heisenberg chain. Specifically, we investigate a long-range one-dimensional interacting Heisenberg spin chain, 
\begin{align}
    H =  \sum_{ij} \frac{1}{r_{ij}^{\delta}} \Le( u_{ij} \hat{\sigma}_i^z \hat{\sigma}_j^z + v_{ij} (\hat{\sigma}_j^{+}\hat{\sigma}_j^{-} + \hat{\sigma}_j^{-}\hat{\sigma}_j^{+}) \Ri)  + \sum_i \varepsilon_i \hat{\sigma}_i^z
\label{eq:long_range_ham}
\end{align} 
where random variables $u_{ij}, v_{ij} \in \pm 1$ and the onsite disorder $\varepsilon_i$ is taken from a box distribution $[-W, W]$. 

This model was analyzed in detail, e.g., in \textcite{BurinPRB15} and in \textcite{Mirlin_longrange2018}. 
Its construction follows the same ideas underlying the power-law random banded matrix model, which has been investigated extensively by \textcite{EversReview08}, \cite{MirlinPRBMPRB00}. Due to its long-range nature, the model \eqref{eq:long_range_ham} exhibits a localization transition characterized by an effective disorder that depends on system size as 
$W_\text{c}  \propto L^{2d - \delta} \ln L$, in the regime when $d < \delta < 2d$.

We proceed to analyze this model from the fragmentation perspective. The results are given in Figure~\ref{fig:long_range} and they summarize as follows: 
\begin{itemize}
\item We first note that the flow of $\overline{\text{\Nc/\Nred}}$ with system size $L$ is faster compared to the short range model, the variance also indicates the same flow. 

\item A similar scaling collapse of $\overline{\text{\Nc/\Nred}}$ with $\mathrm{var}(\text{\Nc/\Nred})$ is observed, as in the short range model shown in Fig.~5(a) of the main text. The small system size data ($L=12$) in Fig.~\ref{fig:long_range}(middle panel) are visibly displaced from the master curve, reflecting strong finite size effects, which are expected in the presence of power law correlations. As $L$ increases, the data flows towards either the ergodic or the strong-disorder fixed point, in close analogy to the short-range model.

\item In the right panel, we show the scaling of $\Wfrag(L)$ with system size. An almost linear scaling is observed of $\Wfrag(L)$, indicating that there is no fragmentation transition in this model. Inset shows the $\mathrm{var}(\text{\Nc/\Nred})$ as a function of rescaled disorder $W/L$, where an approximate collapse is visible.

\end{itemize}

The main point we emphasize is that, even in the presence of power law interactions, a fragmentation transition can be defined for finite $L$. Qualitatively, it exhibits a fragmentation phenomenology similar to that observed in the short-range-interacting model discussed in the main text.

\end{document}